\documentclass[preprint,aps,prc,showpacs,nofootinbib,secnumarabic,floatfix]{revtex4-1}

\usepackage{graphicx}
\usepackage{bm}
\usepackage{color}
\usepackage{gensymb}
\usepackage{amsmath}
\usepackage{slashed}
\usepackage{appendix}

\newcommand{\ltsim}{\protect\raisebox{-0.5ex}{$\:\stackrel{\textstyle <}
	{\sim}\:$}}
\newcommand{\gtsim}{\protect\raisebox{-0.5ex}{$\:\stackrel{\textstyle >}
	{\sim}\:$}}

\newcommand{\bvec}[1]{\ensuremath{\boldsymbol{#1}}}

\begin{document}

\title{Rescattering effects in antiproton-induced exclusive $J/\psi$ and $\psi^\prime$ production on the deuteron}  

\author{A.B. Larionov$^{1,2,3}$\footnote{Corresponding author.\\ 
        E-mail address: larionov@fias.uni-frankfurt.de},
        A. Gillitzer$^1$,M. Strikman$^4$}

\affiliation{$^1$Institut f\"ur Kernphysik, Forschungszentrum J\"ulich, D-52425 J\"ulich, Germany\\
             $^2$National Research Center "Kurchatov Institute", 
             123182 Moscow, Russia\\
             $^3$Frankfurt Institute for Advanced Studies (FIAS), 
             D-60438 Frankfurt am Main, Germany\\
             $^4$Pennsylvania State University, University Park, PA 16802, USA}

\begin{abstract}
  On the basis of the generalized eikonal approximation we study the exclusive reactions $\bar p d \to J/\psi\, n$
  and $\bar p d \to \psi^\prime\, n$ in vicinity of the thresholds for charmonium production on a free proton target.
  It is shown that the rescattering of the incoming antiproton and outgoing charmonium on the spectator neutron
  leads to a depletion of the charmonium production at low- and to an enhancement at high transverse momenta.
  This is in qualitative agreement with previous studies of hard proton knockout in proton-deuteron collisions.
  We analyze different physical sources of uncertainty which may influence the extraction
  of the total charmonium-neutron cross section. The color transparency effect for the incoming $\bar p$ largely
  compensates the influence of charmonium rescattering both at low and high transverse momenta.
  Different choices of the deuteron wave function lead to significant uncertainties at high transverse momenta.
  As an outcome of the calculations of charmonium production, we also provide predictions on the production of
  open charm hadrons due to the dissociation of the charmonium on the neutron. It is shown that the open charm
  production cross section is proportional to the total charmonium-nucleon cross section and quite stable
  with respect to the variation of other parameters of the model. We thus suggest that open charm channels are
  most suited for future studies of charmonium-nucleon interactions at PANDA with a deuteron target.
\end{abstract}
  
\date{\today}

\maketitle

\section{Introduction}
\label{Intro}

The nucleus may serve as a laboratory for studies of interactions of exotic hadrons with nucleons.
In this way, one may test the underlying QCD structures of these particles and explore the dynamics
of transient quark-gluon configurations. Especially interesting are $c \bar c$ states, i.e. charmonia.
Their size is small, and thus at high momenta the interaction cross section with nucleons can be described by the convolution
of the pQCD-calculable color-dipole-nucleon cross section with the $c \bar c$ wave function which effectively
incorporates non-perturbative QCD effects \cite{Gerland:1998bz,Hufner:2000jb}. A small-size $q\bar q$ pair interacts with
the nucleon with a total cross section which is approximately proportional to the squared size of the pair. 
For example, the Cornell potential
gives the r.m.s. distance between $c$ and $\bar c$ of 0.376 fm for $J/\psi(1S)$ and 0.760 fm for $\psi^\prime(2S)$,
while the calculated charmonium-nucleon total cross sections are 3.62 mb for $J/\psi$ and 20.0 mb for $\psi^\prime$
\cite{Gerland:1998bz}. Thus, obtaining the information on the charmonium-nucleon total cross section
would allow to test the size of the charmonium.
Moreover, the knowledge of the charmonium-nucleon dissociation (or absorption) cross sections
is an important ingredient for studies of charmonium suppression in a quark-gluon plasma
\cite{Matsui:1986dk,Vogt:1999cu}.

The phenomenological values of the $J/\psi N$ absorption cross section range from $3.5\pm0.8$ mb determined from the
$A$-dependence of $J/\psi$ photoproduction at $E_\gamma=20$ GeV \cite{Anderson:1976hi} to $6-7$ mb from proton-
and nucleus-nucleus collisions \cite{Gerschel:1993uh,Kharzeev:1996yx}. These analyses rely on a Glauber model description
of $J/\psi$ absorption. For the kinematical conditions of S+Au collisions at $E_{\rm lab}=200$ A GeV, the analysis of
ref. \cite{Abreu:1998wx} concluded the absorption cross sections $\sigma^{\rm abs}_{J/\psi N}=5.68\pm1.92$ mb and
$\sigma^{\rm abs}_{\psi^\prime N} \simeq 4\sigma^{\rm abs}_{J/\psi N}$. However, it is not obvious whether at these high-energy experiments
the produced charmonium states interact with their ``normal'' hadronic cross sections or with reduced ones due to color
transparency (CT) effects. Moreover, at high energies $J/\psi$ mesons may be produced in decays of higher charmonium states.
The possibility of non-diagonal $\psi^\prime N \to J/\psi N$ transitions \cite{Gerland:2005ca} even more complicates
the interpretation of $J/\psi$ production at high energies.

The study of $J/\psi$ production in antiproton-induced reactions on nuclei close to the threshold for the $\bar p p \to J/\psi$
process on a free proton has been suggested to be ideal for the extraction of the $J/\psi N$ cross section
\cite{Brodsky:1988xz,Farrar:1989vr,Larionov:2013axa}.
Due to small size of the charmonium, the multiple gluon exchanges with the target
  nucleon -- ultimately leading to the non-linear QCD behaviour -- are likely to be suppressed even at relatively low energies
  (see, for example, refs. \cite{Frankfurt:2003uj,Gerland:2005ca} where the slope of the momentum dependence of elastic $J/\psi N$ cross section
  has been evaluated using the two-gluon exchange). Thus, the charmonium produced in the $\bar p p \to J/\psi$ channel
 is expected to be slow enough for the formation length effects to be almost negligible,
 but energetic enough for pQCD to be applicable for the description
 of its interaction with a nucleon.   

On the other hand, the information on the $J/\psi N$ cross section at the charmonium momentum of a few GeV/c would be helpful for testing
different theoretical approaches as there are currently no established one for such momenta.
Predictions of different approaches are strongly spread.
The scattering length and range extracted from the lattice QCD calculations \cite{Sugiura:2017vks} give the $J/\psi N$ elastic cross section
growing up to $\simeq 50$ mb at threshold $\sqrt{s} = m_N+m_{J/\psi} = 4.035$ GeV \footnote{We are grateful to J. Haidenbauer for pointing us to this result.}.
Semiphenomenological QCD models \cite{Gerland:1998bz,Hufner:2000jb,Frankfurt:2003uj,Gerland:2005ca}
produce quite slowly increasing total cross section with $\sqrt{s}$.
The parameterization of ref.  \cite{Gerland:2005ca} gives $\sigma_{J/\psi N} = 3.2$ mb at $J/\psi N$ threshold.
In the meson-exchange model \cite{Sibirtsev:2000aw}, the $J/\psi N$ dissociation cross section is calculated as a sum of partial
cross sections producing a peak of $\simeq 2.3$ mb near $\sqrt{s}=4.5$ GeV due to the $\bar D \Lambda_c$ and $\bar D^* \Lambda_c$
contributions on the top of monotonically increasing  $\bar D D N$ contribution. 
The meson-exchange calculations of ref. \cite{Oh:2007ej} are restricted by the $\bar D \Lambda_c$ and $\bar D^* \Lambda_c$ outgoing channels
whose partial contributions are sharply peaked close to respective thresholds and quickly drop at larger $\sqrt{s}$. The values at the peaks
are 2-8 mb for $\bar D \Lambda_c$ and 0.2-2 mb for  $\bar D^* \Lambda_c$, depending on coupling constants and cutoffs.
The coupled-channel calculations of ref. \cite{Molina:2012mv} indicate the dynamically-generated resonance peak of $\simeq 8$ mb in the $\sigma_{J/\psi N}^{\rm inel}$
close to $\sqrt{s}=4.415$ GeV. In the coupled-channel calculations of ref. \cite{Xiao:2015fia} employing the heavy quark spin symmetry,
the total $J/\psi N$ cross section has three dynamically-generated resonance peaks at 4.262 GeV, 4.410 GeV and 4.481 GeV for $J=1/2$
and at 4.334 GeV, 4.417 GeV and 4.481 GeV for $J=3/2$. Remarkably, the peak at 4.481 GeV appears for the both values of the total angular
momentum and is quite close to $\sqrt{s}=4.479$ GeV for the $J/\psi$ produced by a $\bar p p \to J/\psi$ collision.

In this work, we theoretically study $J/\psi$ and $\psi^\prime$ production in $\bar p$-deuteron interactions at beam momenta
close to production thresholds on a free proton ($p_{\rm thr}=4.07$ GeV/c and 6.23 GeV/c, respectively). We also study the open charm
production cross sections due to the charmonium dissociation, i.e. the sum of $\Lambda_c \bar D + X$ and $D\bar D + X$ inclusive cross sections,
which is evaluated as the difference between the integrated charmonium production cross sections without and with charmonium rescattering.
The main purpose is to find kinematical conditions and observables having the largest sensitivity to the charmonium-neutron cross section
\footnote{The $J/\psi N$ and $\psi^\prime N$ elastic cross sections are expected to be on the level of 10\% of the respective
  total cross sections \cite{Gerland:2005ca}. Thus, we do not distinguish between the total and absorption/dissociation
  charmonium-neutron cross sections, although, formally, the total cross section enters in the expression for the
  elastic scattering amplitude, cf. Eq.(\ref{M_psin}) below.}.

The deuteron has the simplest wave function among all complex nuclei and thus serves as a clean testing ground for the theoretical models
that can be extended to heavier targets in the next step. All effects due to particle rescattering, i.e. the initial and final state interactions,
are present for the deuteron target although being reduced as compared to heavier targets.
The reactions $\bar p d \to \Phi n, \Phi=J/\psi, \psi(2S), \psi(3770), \chi_{c2}$ have already been considered in ref. \cite{Cassing:1999wp}
taking into account elastic rescattering of the charmonium on the neutron. The authors of ref.\cite{Cassing:1999wp} concluded that the elastic
$\Phi N$ scattering governs the $\Phi$ production at $p_t > 0.4$ GeV/c. We confirm this conclusion. However, the summation of partial amplitudes
has been performed incoherently and the antiproton elastic rescattering on the neutron prior to the $\bar p p \to \Phi$ transition has been
neglected in ref. \cite{Cassing:1999wp}, which did not allow to address the nuclear absorption. In contrast, we sum the amplitudes coherently
and demonstrate that the antiproton rescattering is important both at small and large transverse momenta of the charmonium leading, respectively,
to a depletion and an enhancement of the charmonium yield. 
In refs.\cite{Cassing:1999wp,Haidenbauer:2008ff} the open charm meson production channels $\bar p d \to D^- D^0 p$ have been considered
above the threshold for the $\bar p n \to D^- D^0$ ($p_{\rm thr}=6.43$ GeV/c) focusing on the effects of the $Dn$ elastic rescattering.
Although we do not specifically address $D\bar D$ production, our results on the open charm production could also serve as a benchmark
for possible future studies of the subthreshold $D \bar D$ production in $\bar p d$ collisions.

The structure of the paper is as follows. The theoretical model is described in sec. \ref{model}. Sec. \ref{obs} includes the formulas
for the differential cross sections and nuclear transparency ratio. Sec. \ref{results} contains the numerical results of calculations
of the differential cross sections of the charmonium production, transverse momentum dependence of the transparency ratio,
and integrated cross sections of charmonium and open charm production. In sec. \ref{discuss}, the expected event rates at PANDA
are estimated. Finally, the summary and conclusions are given in sec. \ref{summary}.

\section{The model}
\label{model}

We employ in our calculations the generalized eikonal approximation to the multiple-scattering theory
\cite{Frankfurt:1994kt,Frankfurt:1996uz,Frankfurt:1996xx,Sargsian:2001ax,Larionov:2013nga}.
Since $J/\psi$ and $\psi^\prime$ have the same quantum numbers $J^{PC}=1^{--}$, all model formalism below is identical for the both charmonia.
Thus, below ``$\psi$'' will denote both $J/\psi$ and $\psi^\prime$ unless specific states are addressed. 

\begin{figure}
\includegraphics[scale = 0.60]{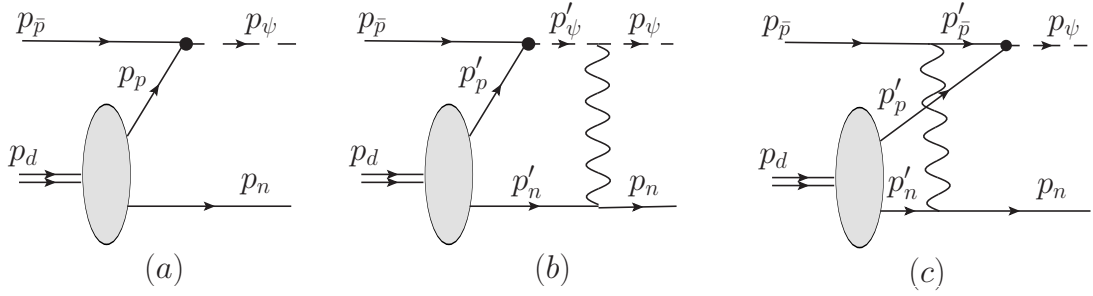}
\caption{\label{fig:graphs} Feynman diagrams for the process $\bar p d \to \psi n$.
The wavy lines denote elastic scattering amplitudes.}
\end{figure}
We take into account the impulse approximation (IA) diagram of Fig.~\ref{fig:graphs}a and
the diagrams with elastic rescattering of $\psi$ and $\bar p$ on the spectator neutron,
Figs.~\ref{fig:graphs}b,c, respectively. The diagrams with $\pi^0$ ($\pi^-$) emission in the
$\psi$ production vertex and absorption on the spectator neutron (proton) should not be added
as they are effectively contained in the IA diagram within the deuteron decay vertex.

The impulse approximation (IA) invariant amplitude of Fig.~\ref{fig:graphs}a is expressed as follows:
\begin{equation}
    M^{(a)}=M_{\psi;\bar pp}(p_\psi,p_{\bar p}) \frac{i\Gamma_{d \to pn}(p_d,p_n)}{p_p^2-m^2+i\epsilon}~,  \label{M^(a)}
\end{equation}
where $M_{\psi;\bar pp}(p_\psi,p_{\bar p})$ is the invariant amplitude of the transition $\bar p p \to \psi$ and $m$ is the nucleon mass.
The deuteron vertex function $\Gamma_{d \to pn}(p_d,p_n)$ is related to the deuteron wave function (DWF) in momentum space
(cf. refs. \cite{Frankfurt:1996uz,Larionov:2018lpk}):
\begin{equation}
  \frac{i \Gamma_{d \to pn}(p_d,p_n)}{p_p^2-m^2+i\epsilon} = \left(\frac{2E_nm_d}{E_p}\right)^{1/2} (2\pi)^{3/2} \phi(\mathbf{p}_p)~,
  \label{Gamma_d2pn}
\end{equation}
where $E_n=\sqrt{\mathbf{p}_n^2+m^2}$ is the neutron energy, $m_d$ is the deuteron mass, $E_p=m_d-E_n$ is the proton energy,
and $\mathbf{p}_p=-\mathbf{p}_n$ is the proton momentum in the deuteron rest frame. Thus, we have
\begin{equation}
    M^{(a)}=M_{\psi;\bar pp}(p_\psi,p_{\bar p}) \left(\frac{2E_nm_d}{E_p}\right)^{1/2} (2\pi)^{3/2} \phi(\mathbf{p}_p)~.  \label{M^(a)_fin}
\end{equation}

The invariant amplitude with $\psi$ rescattering on the neutron (Fig.~\ref{fig:graphs}b) can be written as
\begin{equation}
  M^{(b)} = \int \frac{d^4p_n^\prime}{(2\pi)^4}
  \frac{M_{\psi n}(t)M_{\psi;\bar pp}(p_\psi^\prime,p_{\bar p})\Gamma_{d \to pn}(p_d,p_n^\prime)}%
{(p_\psi^{\prime 2}-m_\psi^2+i\epsilon)(p_n^{\prime 2}-m^2+i\epsilon)(p_p^{\prime 2}-m^2+i\epsilon)}~,    \label{M^(b)}
\end{equation}
where $M_{\psi n}(t)$ is the $\psi n$ elastic scattering amplitude, $t=k^2$ with $k=p_n-p_n^\prime$ being the four-momentum transfer to the neutron.
The integration contour over $dp_n^{\prime 0}$ can be closed in the lower part of the complex plane where only the particle pole of the neutron
propagator contributes. This leads to the following expression:
\begin{equation}
  M^{(b)} = -\int \frac{d^3p_n^\prime}{(2\pi)^3}
  \frac{M_{\psi n}(t)M_{\psi;\bar pp}(p_\psi^\prime,p_{\bar p})}{p_\psi^{\prime 2}-m_\psi^2+i\epsilon}
   \left(\frac{m_d}{2E_n^\prime E_p^\prime}\right)^{1/2} (2\pi)^{3/2} \phi(\mathbf{p}_p^\prime)~,    \label{M^(b)_1}
\end{equation}
where $E_n^\prime=\sqrt{\mathbf{p}_n^{\prime 2}+m^2}$, $E_p^\prime=m_d-E_n^\prime$, $\mathbf{p}_p^\prime=-\mathbf{p}_n^\prime$.
The inverse propagator of $\psi$ can 
be rewritten as (cf. similar case of the knockout nucleon, Eqs.(38),(39),(40) in ref. \cite{Sargsian:2001ax})
\begin{equation}
  p_\psi^{\prime 2}-m_\psi^2+i\epsilon=(p_d-p_n^\prime+p_{\bar p})^2-m_\psi^2+i\epsilon
  =p_d^2 - 2p_dp_n^\prime + p_n^{\prime 2} + 2p_{\bar p}(p_d-p_n^\prime)+ p_{\bar p}^2-m_\psi^2+i\epsilon~.     \label{inv_psi_prop}
\end{equation}
Using the mass-shell condition for the outgoing $\psi$,
\begin{equation}
  m_\psi^2=(p_{\bar p}+p_d-p_n)^2=p_{\bar p}^2+2p_{\bar p}(p_d-p_n)+p_d^2-2p_dp_n+p_n^2~,    \label{psi_on_shell}
\end{equation}
we obtain
\begin{equation}
  p_\psi^{\prime 2}-m_\psi^2+i\epsilon=2p_{\rm lab}(p_n^{\prime z}-p_n^z+\Delta_\psi^0+i\epsilon)~, \label{inv_psi_prop_fin}
\end{equation}
with
\begin{equation}
  \Delta_\psi^0=\frac{(E_{\bar p}+m_d)(E_n-m)}{p_{\rm lab}}~, \label{Delta_psi^0}
\end{equation}
where we neglected the neutron Fermi motion.

Using the coordinate representation of the $\psi$-propagator,
\begin{equation}
   \frac{i}{p_n^{\prime z}-p_n^z+\Delta_\psi^0+i\epsilon}
     =\int dz^0 \Theta(z^0) \mbox{e}^{i(p_n^{\prime z}-p_n^z+\Delta_\psi^0)z^0}~,   \label{Dcoord}
\end{equation}
the relation between the DWFs in momentum and coordinate space,
\begin{equation}
   \phi(\mathbf{p}_p^\prime) = \int \frac{d^3 r}{(2\pi)^{3/2}} \mathbf{e}^{-i\mathbf{p}_p^\prime\mathbf{r}} 
                   \phi(\mathbf{r})~,   \label{phi(p_p)}
\end{equation}
with $\mathbf{r}=\mathbf{r}_p-\mathbf{r}_n$, and setting the intermediate $\psi$ on the mass shell in the elementary
transition matrix element $M_{\psi;\bar pp}$ allows to perform the integration over $dp_n^{\prime z}$ in Eq.(\ref{M^(b)_1})
analytically which gives:
\begin{equation}
  M^{(b)} = \frac{i}{2p_{\rm lab}m^{1/2}} \int d^3r \phi(\mathbf{r}) \Theta(-z) \mbox{e}^{i\mathbf{p}_n\mathbf{r}-i\Delta_\psi^0z}
  \int \frac{d^2k_t}{(2\pi)^2} M_{\psi n}(t) M_{\psi;\bar pp}(p_\psi^\prime,p_{\bar p}) \mbox{e}^{-i\mathbf{k}_t\mathbf{b}}~, \label{M^(b)_2}
\end{equation}
where $t=(E_n-m)^2-k_t^2-(\Delta_\psi^0)^2$.

The comment is in order here with regard to the accuracy of Eq.(\ref{M^(b)_2}).
  The matrix element $M_{\psi;\bar pp}(p_\psi^\prime,p_{\bar p})$ has been set on the $\psi$ mass shell by using
  the four-momentum conservation $p_\psi^\prime=p_{\bar p}+p_d-p_n^\prime$ and an approximate relation $p_n^{\prime z}=p_n^z-\Delta_\psi^0$
  corresponding to the pole of the linearized inverse propagator (\ref{inv_psi_prop_fin}). In doing this we implicitly assumed that the antiproton
  beam momentum is close to the $\psi$ production threshold for free proton target. Under that condition we thus obtain $p_n^{\prime z} \simeq 0$.
  The momentum scale of variation of $M_{\psi;\bar pp}$ is of the order of 1 GeV/c as can be seen from Fig.~\ref{fig:M_ppbarPsi} in sec. \ref{pbarp2psi}.
  Since the $S$-wave component of the deuteron wave function becomes small (on the level of 10\%
  of the maximum value) at momenta above 0.1 GeV/c, practically all uncertainty due to the momentum dependence of the matrix element
  is due to the $D$-wave component which has a probability of $5.8\%$. We have checked that the D-wave total contribution to the rescattering
  amplitudes grows from $\sim 1\%$ to $\sim 10\%$ with increasing transverse momentum of the charmonium from 0 to 0.6 GeV/c.  
  Thus, the estimated accuracy of Eq.(\ref{M^(b)_2}) in calculation of the transverse momentum differential cross section (see Eq.(\ref{dsigma_d2pt}) below)
  is of the order of 1\% at small $p_{\psi t}$ (where the uncertainty is due to interference of IA and rescattering amplitides) and grows to
  $\sim 20\%$ at $p_{\psi t}=0.6$ GeV/c (where  the uncertainty is due to the squared rescattering amplitudes).

Finally, we apply the quasifree approximation in the hard $\bar p p \to \psi$
transition amplitude replacing $p_\psi^\prime \to p_\psi$ which allows to factorize it out.
Then, after integration over azimuthal angle $\phi_{\mbox{k}_t}$ in Eq.(\ref{M^(b)_2}), we come to the following expression:
\begin{equation}
  M^{(b)} = \frac{iM_{\psi;\bar pp}(p_\psi,p_{\bar p})}{4\pi p_{\rm lab}m^{1/2}} \int d^3r \phi(\mathbf{r}) \Theta(-z) \mbox{e}^{i\mathbf{p}_n\mathbf{r}-i\Delta_\psi^0z}
  \int\limits_0^{+\infty} dk_t k_t M_{\psi n}(t) J_0(k_tb)~, \label{M^(b)_fin}
\end{equation}
where $J_0(x)$ is the Bessel function of the first kind.

The invariant amplitude with $\bar p$ rescattering, Fig.~\ref{fig:graphs}c,
is more conveniently written as an integral over four-momentum of the intermediate proton:
\begin{equation}
  M^{(c)} = \int \frac{d^4p_p^\prime}{(2\pi)^4}
  \frac{M_{\bar p n}(t)M_{\psi;\bar pp}(p_\psi,p_{\bar p}^\prime)\Gamma_{d \to pn}(p_d,p_n^\prime)}%
{(p_{\bar p}^{\prime 2}-m^2+i\epsilon)(p_n^{\prime 2}-m^2+i\epsilon)(p_p^{\prime 2}-m^2+i\epsilon)}~,    \label{M^(c)}
\end{equation}
where $M_{\bar p n}(t)$ is the antiproton-neutron elastic scattering amplitude.
Performing the contour integration over $dp_p^{\prime 0}$ in the lower
part of the complex plane, where only the particle pole of the proton propagator contributes, we obtain:
\begin{equation}
  M^{(c)} = -\int \frac{d^3p_n^\prime}{(2\pi)^3}
  \frac{M_{\bar p n}(t)M_{\psi;\bar pp}(p_\psi,p_{\bar p}^\prime)}{p_{\bar p}^{\prime 2}-m^2+i\epsilon}
   \left(\frac{m_d}{2E_n^\prime E_p^\prime}\right)^{1/2} (2\pi)^{3/2} \phi(\mathbf{p}_p^\prime)~,    \label{M^(c)_1}
\end{equation}
with $\bvec{p}_n^\prime=-\bvec{p}_p^\prime$, $E_p^\prime=\sqrt{m^2+\bvec{p}_p^{\prime 2}}$, $E_n^\prime=m_d-E_p^\prime$.
The inverse propagator of the intermediate antiproton can be expressed as
\begin{equation}
  p_{\bar p}^{\prime 2}-m^2+i\epsilon = 2p_{\rm lab}(p_n^z-p_n^{\prime z}-\Delta_{\bar p}^0+i\epsilon)~,  \label{inv_barp_prop_fin}
\end{equation}
where
\begin{equation}
  \Delta_{\bar p}^0=\frac{E_{\bar p}(E_n-E_n^\prime)}{p_{\rm lab}}-\frac{(p_n^\prime-p_n)^2}{2p_{\rm lab}}
  \simeq \frac{(E_{\bar p}+m)(E_n-m)}{p_{\rm lab}}~.      \label{Delta_barp^0}
\end{equation}
Here, in the last step we neglected the neutron Fermi motion. 
By performing the integration over $dp_n^{\prime z}$ in Eq.(\ref{M^(c)_1}) and using Eqs.(\ref{Dcoord}),(\ref{phi(p_p)}) we obtain
the following expression:
\begin{equation}
  M^{(c)} = \frac{i}{2 p_{\rm lab} m^{1/2}} \int d^3r \phi(\mathbf{r}) \Theta(z) \mbox{e}^{i\mathbf{p}_n\mathbf{r}-i\Delta_{\bar p}^0z}
  \int \frac{d^2k_t}{(2\pi)^2} M_{\bar p n}(t) M_{\psi;\bar pp}(p_\psi,p_{\bar p}^\prime) \mbox{e}^{-i\mathbf{k}_t\mathbf{b}}~, \label{M^(c)_2}
\end{equation}
where $t=(E_n-m)^2-k_t^2-(\Delta_{\bar p}^0)^2$.
Finally, we again apply the quasifree approximation in the hard $\bar p p \to \psi$ transition amplitude by replacing
$p_{\bar p}^\prime \to p_{\bar p}$ in it and perform the integration over $\phi_{\mbox{k}_t}$ in Eq.(\ref{M^(c)_2}) which gives:
\begin{equation}
  M^{(c)} = \frac{iM_{\psi;\bar pp}(p_\psi,p_{\bar p})}{4\pi p_{\rm lab} m^{1/2}} \int d^3r \phi(\mathbf{r}) \Theta(z) \mbox{e}^{i\mathbf{p}_n\mathbf{r}-i\Delta_{\bar p}^0z}
  \int\limits_0^{+\infty} dk_t k_t M_{\bar p n}(t) J_0(k_t b)~. \label{M^(c)_fin}
\end{equation}

Note that $\Delta_\psi^0=\Delta_{\bar p}^0=E_n-m$ in the limit of high antiproton energy. Thus, the propagators of the intermediate $\psi$ and $\bar p$
have poles for $p_n^{\prime -} = p_n^-$, where $p^- \equiv p^0 - p^z$. This reflects the conservation of the $p_n^-$ component of the neutron four-momentum
which is a generic feature of high-energy elastic scattering \cite{Frankfurt:1994kt,Sargsian:2001ax}.

We used the coordinate space wave functions in order to include CT effects as it will be discussed below in sec. \ref{CT}.
However, without the CT effects, it is also possible to perform the integration over $dp_n^{\prime z}$ in Eqs.(\ref{M^(b)_1}) and (\ref{M^(c)_1})
with the linearized inverse propagators (\ref{inv_psi_prop_fin}),(\ref{inv_barp_prop_fin}) by closing the integration contour, respectively, in the upper
and lower parts of the complex plane and taking the residues at the poles of the DWF (cf. Appendix A in ref. \cite{Frankfurt:1996uz}).
This leads to the following expressions for the rescattering amplitudes:
\begin{eqnarray}
  M^{(b)} &=& -\frac{M_{\psi;\bar pp}(p_\psi,p_{\bar p})}{2^{7/2} \pi p_{\rm lab}m^{1/2}} \int d^2p_{nt}^\prime M_{\psi n}(t)
               \sum_j \frac{\phi_j^M(-im_{jt},-\mathbf{p}_{nt}^\prime)}{(im_{jt}-p_n^z+\Delta_\psi^0)m_{jt}}~,         \label{M^(b)_MomSpace}\\
  M^{(c)} &=& -\frac{M_{\psi;\bar pp}(p_\psi,p_{\bar p})}{2^{7/2} \pi p_{\rm lab}m^{1/2}} \int d^2p_{nt}^\prime M_{\bar p n}(t)
               \sum_j \frac{\phi_j^M(im_{jt},-\mathbf{p}_{nt}^\prime)}{(im_{jt}+p_n^z-\Delta_{\bar p}^0)m_{jt}}~,      \label{M^(c)_MomSpace}
\end{eqnarray}
where $m_{jt} = \sqrt{m_j^2+\mathbf{p}_{nt}^{\prime 2}}$.
Here, we used the parameterization of the DWF originally proposed for the Paris potential model \cite{Lacombe:1981eg}:
\begin{equation}
     \phi(\mathbf{q})=\frac{1}{\sqrt{4\pi}} \sum_j \frac{\phi_j^M(\mathbf{q})}{q^2+m_j^2}~,      \label{phi_decomp}
\end{equation}
where
\begin{equation}
    \phi_j^M(\mathbf{q}) = \left(\frac{2}{\pi}\right)^{1/2} \left(c_j+\frac{d_j}{\sqrt{8}}S(\mathbf{q})\right) \chi^M~,     \label{phi_j^M}
\end{equation}
with the spin tensor operator
\begin{equation}
   S(\mathbf{q})=\frac{3(\bvec{\sigma}_p\mathbf{q})(\bvec{\sigma}_n\mathbf{q})}{q^2} 
               -\bvec{\sigma}_p\bvec{\sigma}_n~,   \label{Sspin}
\end{equation}
and $\chi^M$ being the eigenfunction of the spin~=~1 state with spin projection $M=0,\pm1$.

In numerical calculations, Eqs.(\ref{M^(b)_MomSpace}),(\ref{M^(c)_MomSpace}) produce indistinguishable results
as compared to Eqs.(\ref{M^(b)_fin}),(\ref{M^(c)_fin}), respectively. However, without CT effects, the numerical calculations
are much faster with Eqs.(\ref{M^(b)_MomSpace}),(\ref{M^(c)_MomSpace}). This allowed us also to check the validity of the
quasifree approximation, which turned out to be working quite well (accuracy of $\sim 3-15\%$ in the transparency ratio at
large transverse momenta, cf. solid and dotted lines on Figs.~\ref{fig:T}a,\ref{fig:T_Psip}a below).

\subsection{Elementary amplitudes}
\label{elem}

\subsubsection{$\bar p p \to \psi$}
\label{pbarp2psi}

In order to describe the $\bar p p \to \psi$ transition amplitude, we will apply the effective Lagrangian with couplings of the Dirac ($\gamma_\mu$)
and Pauli ($\sigma_{\mu\nu}$) type \cite{Barnes:2007ub}:
\begin{equation}
  {\cal L}_{\psi NN} = -g \bar N(\gamma_\mu  - \frac{\kappa}{2m} \sigma_{\mu\nu} \partial^\nu_\psi)N \psi^\mu~,  \label{L_psiNN}
\end{equation}
where $N$ is the nucleon field, $\psi^\mu$ is the $\psi$ field,
and $\sigma_{\mu\nu} = \frac{i}{2}(\gamma_\mu\gamma_\nu-\gamma_\nu\gamma_\mu)$.
The invariant amplitude of the $\bar p p \to \psi$ transition
is thus expressed as
\begin{equation}
  M_{\psi;\bar pp}(p_\psi,p_{\bar p}) = -g\bar u(-p_{\bar p},-\lambda_{\bar p}) (\gamma_\mu  - \frac{i\kappa}{2m} \sigma_{\mu\nu} p_\psi^\nu) u(p_p,\lambda_p) \varepsilon^{(\lambda)\mu *}~, 
    \label{M_psi;barpp}
\end{equation}
with $p_\psi=p_{\bar p}+p_p$ and $\varepsilon^{(\lambda)\mu}$  being the $\psi$ four-momentum and polarization four-vector, respectively. They satisfy the orthogonality
and normalization conditions $p_\psi \varepsilon^{(\lambda)}=0$ and $\varepsilon^{(\lambda^\prime) *} \varepsilon^{(\lambda)} =-\delta_{\lambda^\prime\lambda}$.
For the Dirac spinors, the orthogonality and normalization conditions are 
$\bar u(p_p,\lambda_p^\prime) u(p_p,\lambda_p) =2m\delta_{\lambda_p^\prime\lambda_p}$
and $\bar u(-p_{\bar p},-\lambda_{\bar p}^\prime) u(-p_{\bar p},-\lambda_{\bar p}) =-2m\delta_{\lambda_{\bar p}^\prime\lambda_{\bar p}}$.
We use the spin basis wave functions throughout the paper.
Thus, $\lambda_{\bar p}=\pm1/2$, $\lambda_p=\pm1/2$ and $\lambda=0,\pm1$ denote the spin projections on the $z$-axis
(along the $\bar p$ beam direction) for the antiproton, proton and $\psi$, respectively.

The coupling constant $\kappa$ is fixed by the condition that the angular distribution of $e^+ e^- \to \psi \to p \bar p$ scattering
in the center-of-mass (c.m.) frame is $d\sigma/d\cos(\Theta_{c.m.}) \propto 1+\alpha \cos^2(\Theta_{c.m.})$ where the parameter $\alpha$ is expressed as
(see ref. \cite{Barnes:2007ub})
\begin{equation}
  \alpha=\frac{1-\frac{4m^2}{m_\psi^2}\left|\frac{G_E}{G_M}\right|^2}{1+\frac{4m^2}{m_\psi^2}\left|\frac{G_E}{G_M}\right|^2}~, \label{alpha}
\end{equation}
with $m_\psi$ being the charmonium mass, and $G_E=g(1+\kappa m_\psi^2/4m^2)$ and $G_M=g(1+\kappa)$ being the Sachs-type formfactors. 
The coupling constant $g$ is fixed by the total charmonium width $\Gamma$ and the branching
fraction ${\cal B}(\psi \to \bar p p)$ as expressed by
\begin{equation}
  \Gamma \cdot {\cal B}(\psi \to \bar p p)=\frac{2m^2|G_E|^2+m_\psi^2|G_M|^2}{12\pi m_\psi} \sqrt{1-\frac{4m^2}{m_\psi^2}}~. \label{Gamma_jpsi_to_pbarp}
\end{equation}
The input phenomenological parameters and the determined values of the coupling constants are collected in Table~\ref{tab:pbarp2psi}. 
\begin{table}[htb]
  \caption{\label{tab:pbarp2psi} The coupling constants $\kappa$ and $g$ of the effective Lagrangian (\ref{L_psiNN}), 
    total width $\Gamma$, branching fraction ${\cal B}(\psi \to \bar p p)$, and the anisotropy parameter $\alpha$
    for $J/\psi$ and $\psi^\prime$. The value of $\kappa$ for $\psi^\prime$ is obtained by setting $\alpha=1$.}
  \begin{center}
    \begin{tabular}{|c|c|c|c|c|c|c|}
    \hline
                       & $\kappa$ & $g$                  & $\Gamma$ (keV) & ${\cal B}(\psi \to \bar p p)$  & $\alpha$                 & refs. \\
    \hline
    $J/\psi$           & -0.089   & $1.79 \cdot 10^{-3}$  & $92.9\pm2.8$   & $(2.121\pm0.029) \cdot 10^{-3}$ & $0.595\pm0.012\pm0.015$  & \cite{Ablikim:2012eu,Tanabashi:2018oca}\\
    $\psi^\prime$       & -0.259   & $1.35 \cdot 10^{-3}$  & $294\pm8$      & $(2.88\pm0.10) \cdot 10^{-4}$   & $1.03\pm0.06\pm0.03$     & \cite{Ablikim:2018ttl,Tanabashi:2018oca}\\
    \hline
    \end{tabular}
  \end{center}
\end{table}

\begin{figure}
  \includegraphics[scale = 0.53]{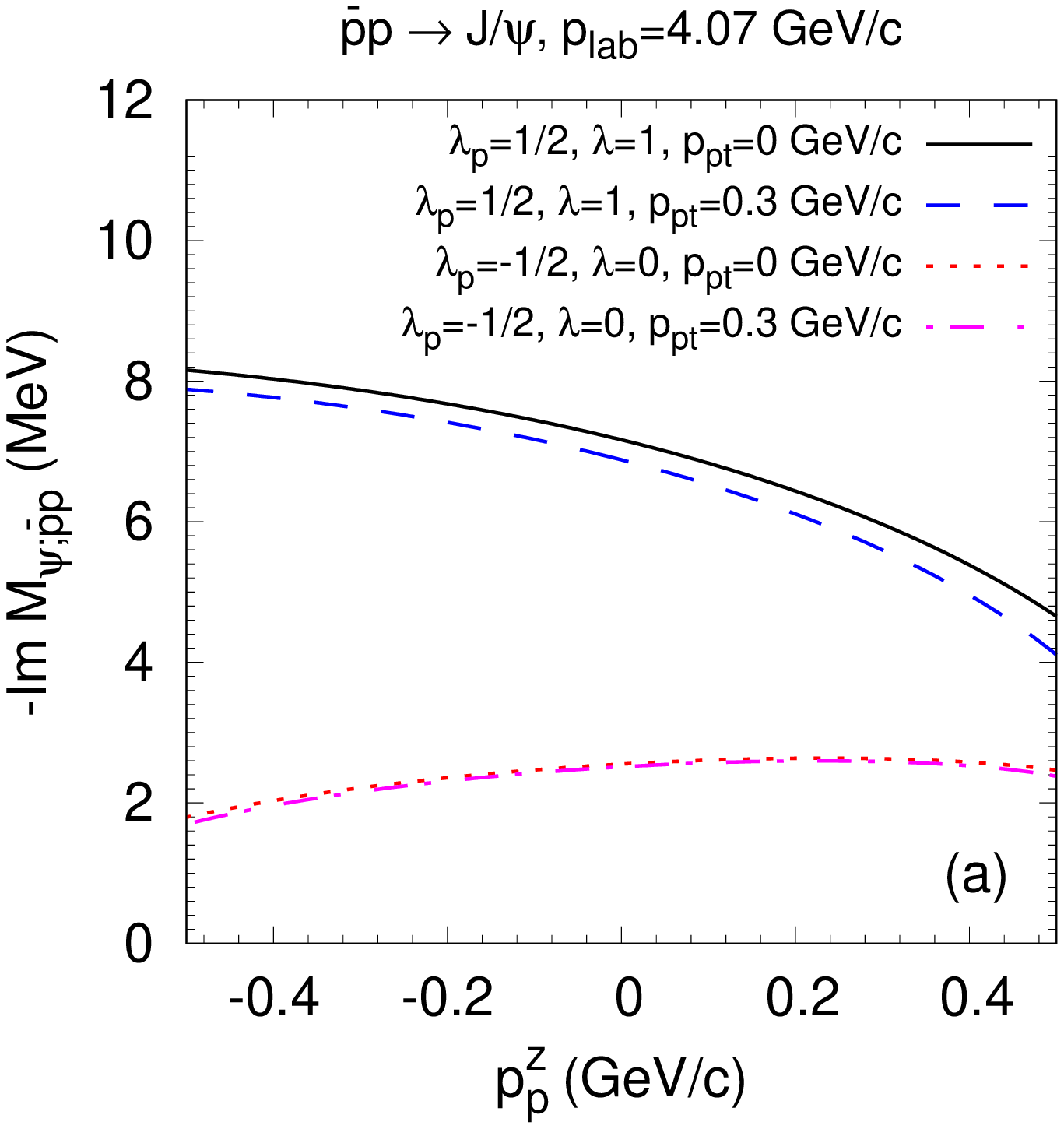}
  \includegraphics[scale = 0.53]{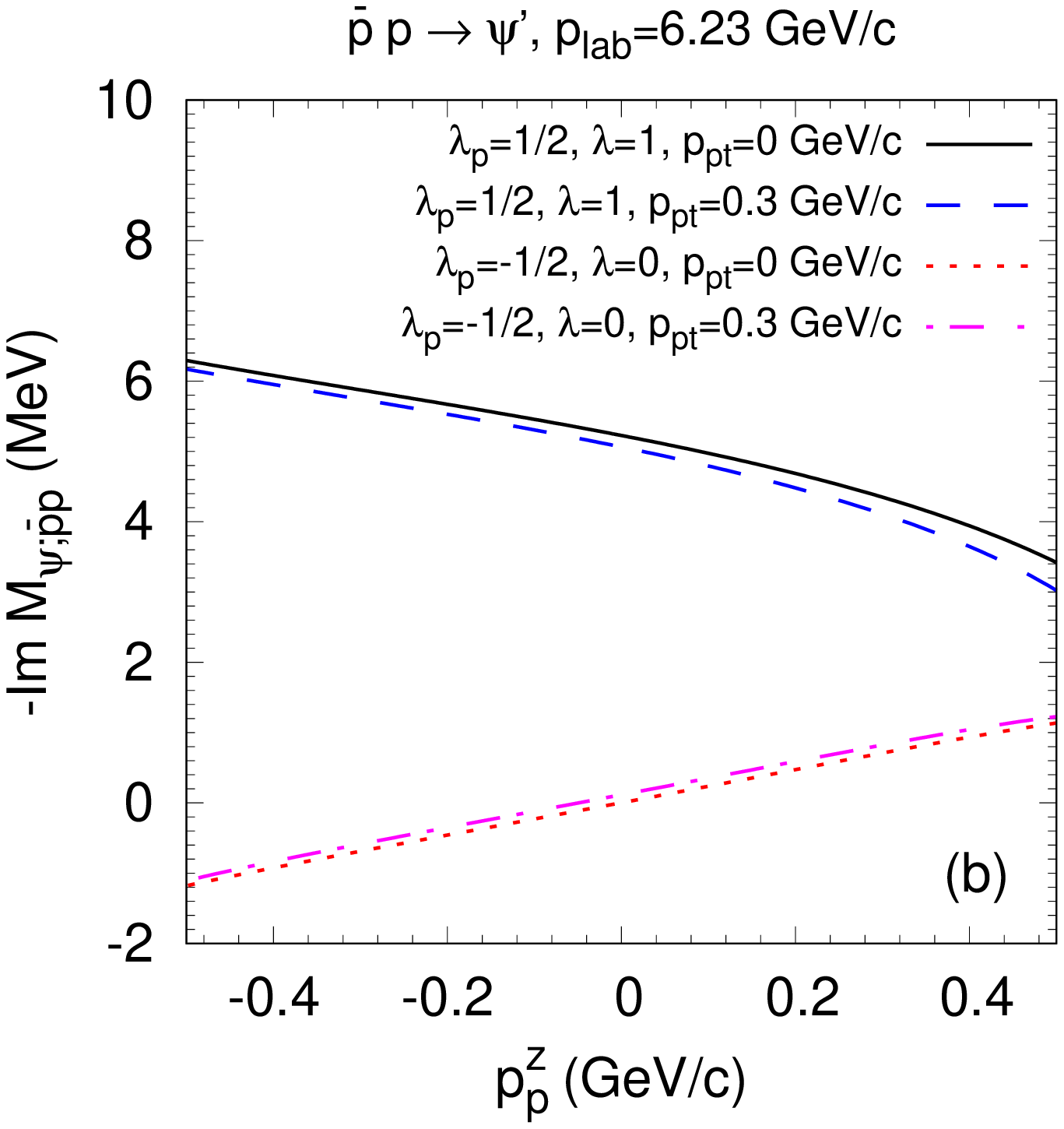}
  \caption{\label{fig:M_ppbarPsi} Imaginary part of the matrix element of the $\bar p p \to J/\psi$ (a) and $\bar p p \to \psi^\prime$ (b)
    transitions as a function of the proton longitudinal momentum for the different spin projections of the proton and charmonium and transverse momenta
    of the proton (along $x$-direction), as indicated. The spin projection of the antiproton fixed to +1/2.
    The antiproton momentum is set to the respective threshold values for free proton target at rest.}
\end{figure}
Figs.~\ref{fig:M_ppbarPsi}a,b show the imaginary part of matrix element, Eq.(\ref{M_psi;barpp}), plotted vs $z$-component of proton momentum
\footnote{The real part of the matrix element of Eq.(\ref{M_psi;barpp}) is equal to zero for wave functions in spin basis.}. The proton energy
is determined as $E_p=m_d-\sqrt{\bvec{p}_p^2+m^2}$, i.e. according to the kinematics of the amplitude with charmonium rescattering (Fig.~\ref{fig:graphs}b).
At $p_p^z=0$ and $p_{pt}=0$  one can determine the ratio (cf. ref. \cite{Olsson:1986aw})
\begin{equation}
  {\cal R} = \frac{2|B_1|^2}{|B_0|^2+2|B_1|^2}
           = \frac{|M_{\psi;\bar pp}(\lambda_p=1/2,\lambda_{\bar p}=1/2,\lambda=1)|^2}%
                  {|M_{\psi;\bar pp}(-1/2,1/2,0)|^2 + |M_{\psi;\bar pp}(1/2,1/2,1)|^2}~,     \label{calR}
\end{equation}
where $|B_0|^2$ and $2|B_1|^2$ are the probabilities to produce $\psi$ with helicities 0 and $\pm1$, respectively, in an unpolarized $\bar p p$ collision,
with a normalization condition being $|B_0|^2+2|B_1|^2=1$. In the last step of Eq.(\ref{calR}), we used the parity relation for helicity amplitudes
(see Eq.(13) in ref. \cite{Ridener:1994ij}).
The values of the anisotropy parameter $\alpha$ in Table \ref{tab:pbarp2psi} can be also obtained from the following relation
(cf. ref. \cite{Olsson:1986aw}):
\begin{equation}
  \alpha=\frac{3{\cal R}-2}{2-{\cal R}}~,    \label{alpha_via_R}
\end{equation}
where ${\cal R}=0.887$ for $J/\psi$ and ${\cal R}=1$ for $\psi^\prime$.

\subsubsection{$\psi n \to \psi n$}
\label{psin_to_psin}

We assume that elastic scattering does not change the spin projections of the scattered
charmonium and neutron. For small momentum transfers, where the spin flip amplitude is known to be small,
this approximates the conservation of particle helicities in the $\psi n$ c.m. frame,
as the spin quantization axis is chosen along the $\bar p$ beam momentum\footnote{The results for the differential cross sections
  in the quasifree approximation applied in our calculations are anyway insensitive to either helicity- or spin projection conservation in elastic rescattering,
  because in this approximation the DWF is projected on the spinor of final state neutron.}.

The amplitude of charmonium elastic scattering on the neutron is assumed to be spin-independent and can be expressed as   
\begin{equation}
    M_{\psi n}(t) = 2 i p_{\rm lab} m \sigma_{\psi n}^{\rm tot}
                    (1-i\rho_{\psi n}) \mbox{e}^{B_{\psi n}t/2}~.                  \label{M_psin}
\end{equation}
Here $\rho_{\psi n}=\mbox{Re}M_{\psi n}(0)/\mbox{Im}M_{\psi n}(0)$. Phenomenological information on the parameters
$\sigma_{\psi n}^{\rm tot}$, $B_{\psi n}$ and $\rho_{\psi n}$ is quite scarce. The ranges of the values for the total charmonium-nucleon
cross sections have been already discussed in sec. \ref{Intro}. The slope of the momentum dependence can be estimated on the basis
of the two-gluon exchange calculation \cite{Gerland:2005ca} which leads to $B_{\psi n} \simeq 3$ GeV$^{-2}$.
The ratio of the real-to-imaginary part of the forward scattering amplitude $\rho_{\psi n}$ is most likely  within the interval
$0.15-0.3$ (cf. ref. \cite{Larionov:2013nga}). If not specified otherwise, we apply the following values
of the parameters of the charmonium-neutron scattering amplitude in the calculations:
$\sigma_{J/\psi n}^{\rm tot}=4$ mb, $\sigma_{\psi^\prime n}^{\rm tot}=20$ mb, $B_{J/\psi n}=B_{\psi^\prime n}=3$ GeV$^{-2}$,
and $\rho_{J/\psi n}=\rho_{\psi^\prime n}=0.2$.

\subsubsection{$\bar p n \to \bar p n$}
\label{barpn_to_barpn}

The invariant amplitude of antiproton-neutron elastic scattering is also assumed to conserve particle helicities and described
by the expression
\begin{equation}
    M_{\bar p n}(t)= 2 i p_{\rm lab} m \sigma_{\bar p n}^{\rm tot}
     (1-i\rho_{\bar p n}) \mbox{e}^{B_{\bar p n}t/2}~,           \label{M_pbarn}
\end{equation}
The parameters $\sigma_{\bar p n}^{\rm tot}$ and $B_{\bar p n}$ are chosen to be equal to the corresponding
$\bar p p$ parameters according to ref. \cite{Larionov:2016xeb}. However, in the present work, we fix $\rho_{\bar p n}=0.05$
which better agrees with phenomenological values of $\rho_{\bar p p}$ at beam momenta of 4-6 GeV/c \cite{Tanabashi:2018oca}.
As in the case of $\psi n$ elastic scattering, we use the simplification that the spin projections on the beam axis are not changed
in $\bar p n$ elastic scattering.

\subsection{Color transparency effects}
\label{CT}

We intend to include CT effects on the amplitude level in the spirit of ref. \cite{Frankfurt:1994kt}.
Point-like configurations (PLCs) participating in a hard collision interact with nucleons with reduced strength.
Since a PLC is squeezed in coordinate space, its formfactor in the momentum space expands. Thus, the slope
of the momentum-transfer dependence of the elastic scattering amplitude of a PLC on a nucleon gets less steep.
However, at moderate energies, which are the subject of our present study, the PLC expands and reaches the normal hadronic size
within the nuclear target already. The latter effect can be accounted for within the quantum diffusion model (QDM) \cite{Farrar:1988me}.

According to the QDM, the effective interaction cross section of a charmonium with a nucleon is expressed as
\begin{equation}
   \sigma_{\psi N}^{\rm eff}(p_{\psi},|z|)
  =\sigma_{\psi N}^{\rm tot} \left(\left[ \frac{|z|}{l_{\psi}}
    + \frac{\langle n_{\psi}^2k_{\psi t}^2\rangle}{m^2_{\psi}} \left(1-\frac{|z|}{l_{\psi}}\right) \right]
    \Theta(l_{\psi}-|z|) +\Theta(|z|-l_{\psi})\right)~,          \label{sigma_PsiN_eff}
\end{equation}
where $|z|$ is the distance traveled by $\psi$ from the hard $\bar p p \to \psi$ interaction point. The charmonium coherence lengths can
be estimated as
\begin{eqnarray}
  l_{J/\psi} &\simeq& \frac{2p_{J/\psi}}{m_{\psi^\prime}^2-m_{J/\psi}^2}~,                         \label{l_Jpsi}\\
  l_{\psi^\prime} &\simeq& 2 l_{J/\psi} \simeq 6\,\mbox{fm}\frac{p_{\psi^\prime}}{30\,\mbox{GeV}}~,      \label{l_PsiPrime}
\end{eqnarray}
as suggested in ref. \cite{Gerland:1998bz}.
The initial squeezing factor $\langle n_{\psi}^2k_{\psi t}^2 \rangle/m^2_{\psi}$ depends on the internal transverse momentum scale
$\sqrt{\langle k_{\psi t}^2\rangle}$ of the charmonium and on the number of valence quarks $n_{\psi}=2$.
In the calculations, we apply the values $\sqrt{\langle k_{J/\psi t}^2\rangle}=0.8$ GeV/c and $\sqrt{\langle k_{\psi^\prime\,t}^2\rangle}=0.4$ GeV/c,
since the size of the $\psi^\prime$ state is about two times the size of the $J/\psi$.
CT effects on the charmonium propagation are introduced by replacing
$\sigma_{\psi n}^{\rm tot} \to \sigma_{\psi N}^{\rm eff}(p_{\psi},|z|)$ in Eq.(\ref{M_psin}).
For simplicity, we will disregard CT effects on the slope $B_{\psi n}$, since its value is not fixed by any experimental data.

CT effects on the $\bar p$ propagation before hard $\bar p p \to \psi$ interaction are introduced in the following way.
The total $\bar p n$ cross section $\sigma_{\bar p n}^{\rm tot}$ in Eq.(\ref{M_pbarn}) is replaced by the effective one:
\begin{equation}
   \sigma_{\bar p N}^{\rm eff}(p_{\bar p},|z|)
  = \sigma_{\bar p N}^{\rm tot}\left(\left[ \frac{|z|}{l_{\bar p}}
    + \frac{\langle n_{\bar p}^2k_{\bar p t}^2\rangle}{m^2_{\psi}} \left(1-\frac{|z|}{l_{\bar p}}\right) \right]
    \Theta(l_{\bar p}-|z|) +\Theta(z-l_{\bar p})\right)~,           \label{sigma_barpn_eff}
\end{equation}
where the antiproton coherence length is
\begin{equation}
   l_{\bar p} = \frac{2p_{\bar p}}{\Delta M^2}~,                     \label{l_barp}
\end{equation}
with $\Delta M^2=0.7$ GeV$^2$, $n_{\bar p}=3$, and $\sqrt{\langle k_{\bar p t}^2\rangle}=0.35$ GeV/c.

In the elastic $\bar p n$ scattering amplitude, only the antiproton enters as a PLC, and hence only the formfactor from
the $\bar p$ side of the amplitude gets modified due to CT. This effect can be described by multiplying the
$\bar p n$ scattering amplitude by the ratio
\begin{equation}
  R = \frac{G_N(t \cdot \sigma_{\bar p N}^{\rm eff}(p_{\bar p},|z|)/\sigma_{\bar p N}^{\rm tot})}{G_N(t)}~,    \label{R}
\end{equation}
where $G_N(t)=1/(1-t/0.71\,\mbox{GeV}^2)^2$ is the Sachs electric formfactor of the proton in the dipole approximation.

\section{Observables}
\label{obs}

The invariant differential cross section of $\bar p d \to \psi n$ process is expressed as
\begin{equation}
  d\sigma_{\bar p d \to \psi n} = (2\pi)^4 \delta^{(4)}(p_{\bar p}+p_d-p_\psi-p_n) \frac{|M^{(a)}+M^{(b)}+M^{(c)}|^2}{4I}
  \frac{d^3p_\psi}{(2\pi)^32E_\psi} \frac{d^3p_n}{(2\pi)^32E_n}~,       \label{dsigma}
\end{equation}
where $I=((p_{\bar p}p_d)^2-m^2m_d^2)^{1/2}=p_{\rm lab} m_d$ is the flux factor. In the deuteron rest frame, the transverse momentum
differential cross section of exclusive $\psi$ production can be calculated as follows:
\begin{equation}
  \frac{d^2\sigma_{\bar p d \to \psi n}}{d^2p_{\psi t}}
  = \frac{|M^{(a)}+M^{(b)}+M^{(c)}|^2}{|p_\psi^z(E_{\bar p}+m_d)-p_{\rm lab}E_\psi| 64\pi^2 p_{\rm lab} m_d}~.    \label{dsigma_d2pt}
\end{equation}
To obtain Eq.(\ref{dsigma_d2pt}), we neglected backward charmonium production in the $\bar p d$ c.m. frame,
since this process is strongly suppressed by the DWF in the IA calculation, and by the elastic scattering amplitudes
in the calculation beyond IA. This results in a unique value of the longitudinal momentum $p_\psi^z$ for a given
transverse momentum $p_{\psi t}$. In actual numerical calculations, the region of charmonium transverse momenta, where the
cross section is nonzero, is restricted by the condition of a time-like struck proton in the IA amplitude\footnote{Otherwise it
  is impossible to construct the Dirac spinor of the struck proton in the matrix element of the $\bar p p \to \psi$ transition,
  Eq.(\ref{M_psi;barpp}).}. 

In order to characterize the longitudinal motion of the charmonium, it is convenient to use the (longitudinal boost invariant)
light cone momentum fraction of the colliding antiproton and proton at rest carried by the charmonium:
\begin{equation}
    \alpha_\psi=\frac{E_\psi+p_\psi^z}{E_{\bar p}+m+p_{\rm lab}}~.       \label{alpha_psi}
\end{equation}
The differential cross section in the variable $\alpha_\psi$ is expressed as:
\begin{equation}
  \frac{d\sigma_{\bar p d \to \psi n}}{d\alpha_\psi}
  = \frac{|M^{(a)}+M^{(b)}+M^{(c)}|^2}{32\pi p_{\rm lab} m_d} \frac{E_{\bar p}+m+p_{\rm lab}}{E_{\bar p}+m_d+p_{\rm lab}}~.    \label{dsigma_dalpha_psi}
\end{equation}

To better see the effects of rescattering we also introduce the transparency ratio as
\begin{equation}
   T=\frac{|M^{(a)}+M^{(b)}+M^{(c)}|^2}{|M^{(a)}|^2}~.      \label{T}
\end{equation}
Thus, the transparency ratio is identically equal to 1 for the IA calculation, i.e. when both $M^{(b)}$ and $M^{(c)}$ are set to zero.
Below, for brevity, we will call the calculation with $M^{(b)}$ set to zero as the calculation 'with $\bar p$ rescattering',
and the full calculation -- as the calculation 'with $\bar p$ and $J/\psi$($\psi^\prime$) rescattering'.

\section{Numerical results}
\label{results}

The DWF is the most important ingredient in our calculations.
It has been calculated in various approaches, e.g. Paris potential \cite{Lacombe:1980dr,Lacombe:1981eg} ($P_D=5.8\%$),
Bonn potential \cite{Machleidt:1987hj} ($P_D=4.3\%$), coupled-channel folded diagram potential \cite{Haidenbauer:1993pw} ($P_D=5.6\%$),
Argonne V18 potential \cite{Wiringa:1994wb} ($P_D=5.8\%$), Nijmegen potential \cite{Stoks:1994wp} ($P_D=5.8\%$),
CD-Bonn potential \cite{Machleidt:2000ge} ($P_D=4.9\%$). Here the fraction of the $D$-wave is given in parentheses.
All potentials predict a very similar wave function in the low-momentum region, however, a significant discrepancy exists
at $p > 0.4$ GeV/c (c.f. Fig.~2 in ref. \cite{Gilman:2001yh}), mostly due to the $D$-wave component.
Since most of the potentials give $P_D=5.6-5.8\%$ for the fraction of the $D$-wave, we will in default calculations
use the DWF of the Paris potential, which can be considered as a conservative choice.
In some selected cases, in order to see the sensitivity to the high-momentum part of the DWF,
we will also display the results obtained with the CD-Bonn potential. 

\begin{figure}
  \includegraphics[scale = 0.53]{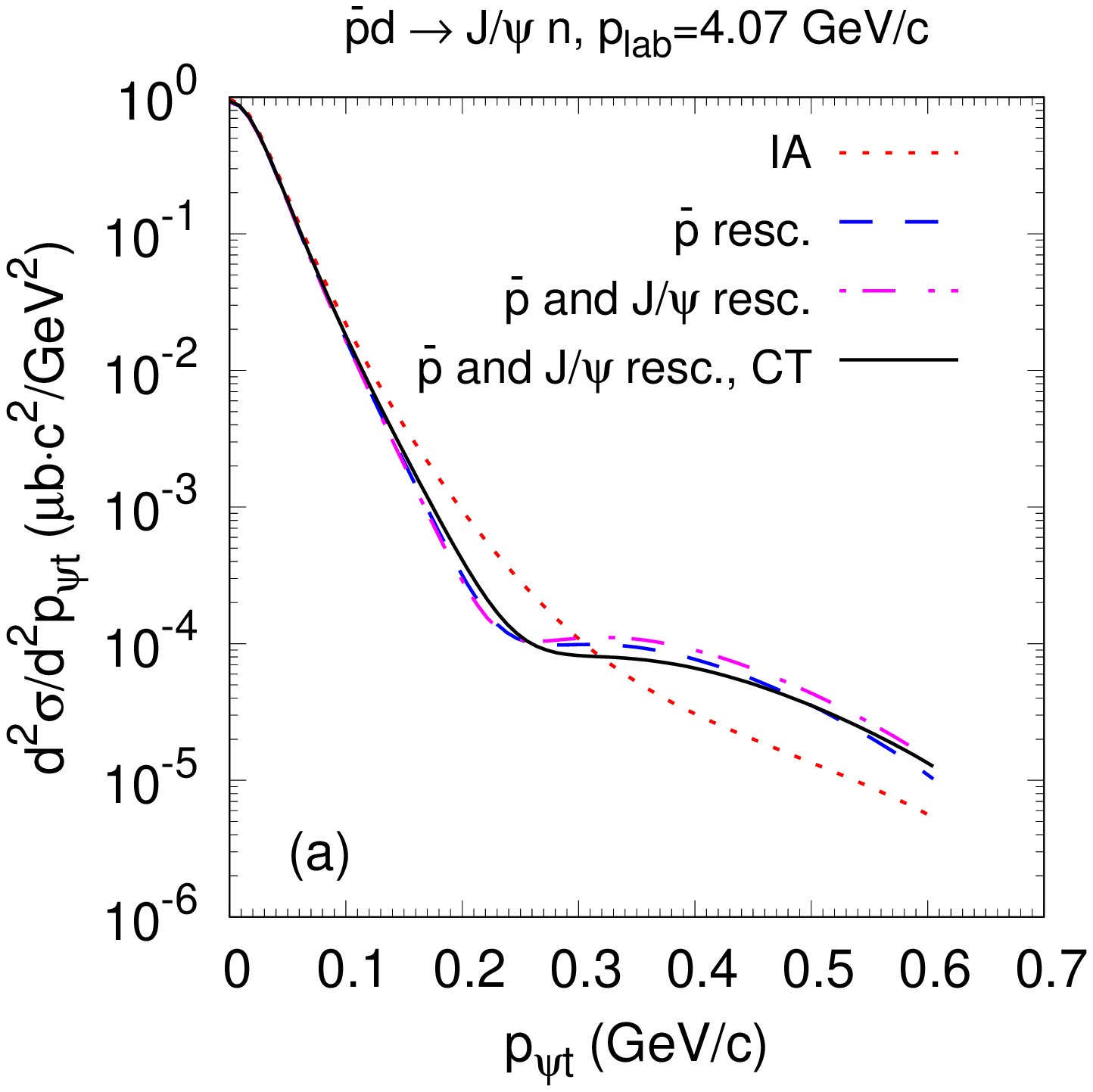}
  \includegraphics[scale = 0.53]{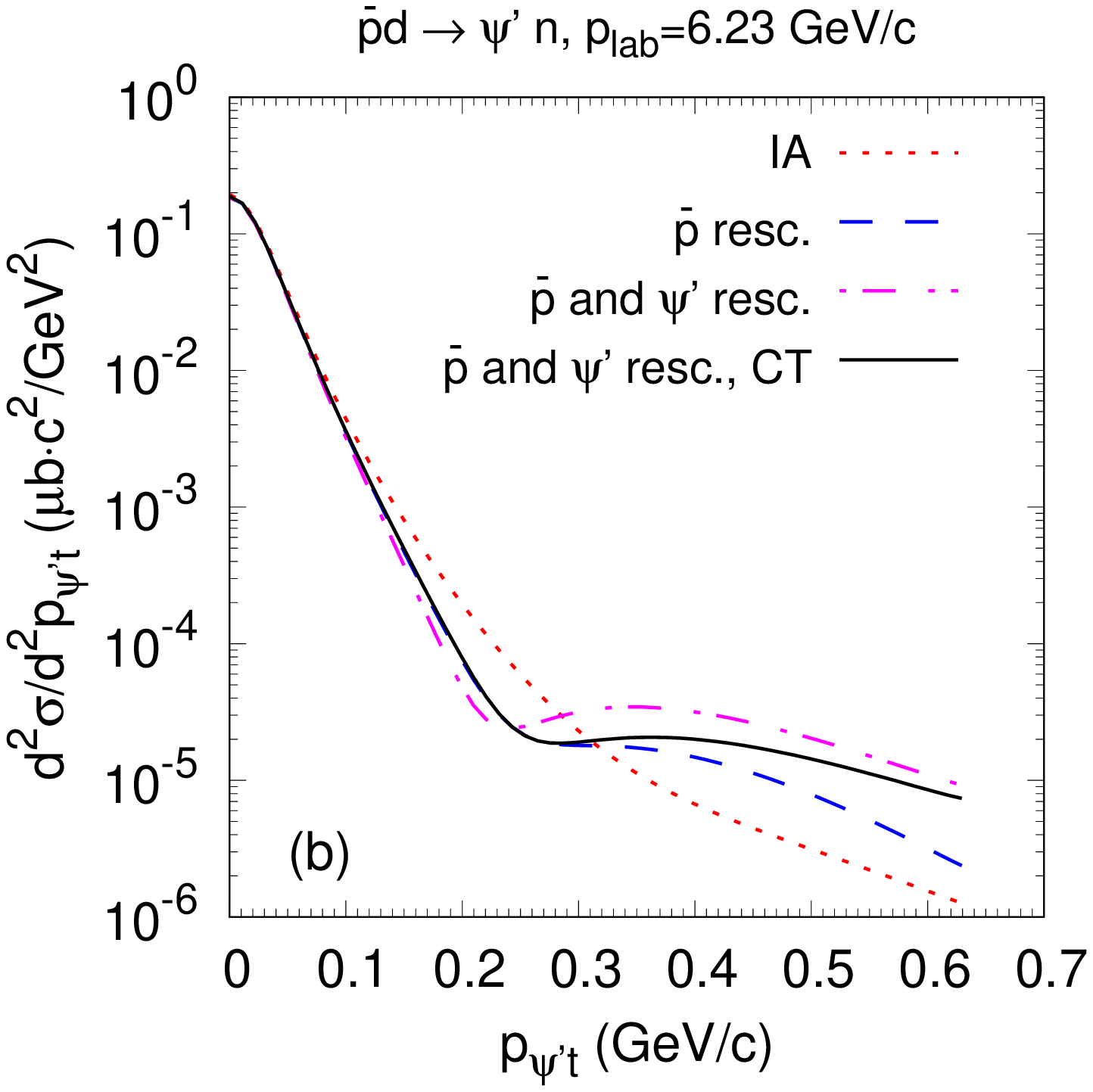}
  \caption{\label{fig:dsigd2pt} Transverse momentum differential cross section of charmonium production in the $\bar p d \to J/\psi\, n$ process
    at 4.07 GeV/c (a) and in the $\bar p d \to \psi^\prime\, n$ process at 6.23 GeV/c (b). The beam momenta correspond to the thresholds for
    charmonium production on the proton at rest.
    Dotted line -- IA, dashed line -- $\bar p$ rescattering without CT, dash-dotted line -- $\bar p$ and $J/\psi$ rescattering without CT,  
    solid line -- $\bar p$ and $J/\psi$ rescattering with CT.}
\end{figure}
Fig.~\ref{fig:dsigd2pt}a shows the transverse momentum differential cross section of $J/\psi$ production at $p_{\rm lab}=4.07$ GeV/c.
In the case of IA, the cross section drops quickly with $p_{\psi t}$ due to the DWF being quite narrow in momentum space.
The effect of the $\bar p$ rescattering is the reduction of the cross section at small and an enhancement at large transverse
momenta of the charmonium.
This is expected, since the elastic rescattering should increase the transverse momentum of a scattered particle.
$J/\psi$ rescattering influences the transverse momentum differential cross section of $J/\psi$ production in a similar way,
but much more weakly, since the $J/\psi\, n$ cross section is quite small.
The introduction of CT partly compensates the effect of rescattering.
We have checked that the CT effect is almost entirely caused by the reduced $\bar p n$ cross section within the antiproton coherence length,
while the reduced $J/\psi\, n$ cross section practically does not influence the results due to the smallness of the $J/\psi$ coherence length.

As shown in Fig.~\ref{fig:dsigd2pt}b, in the case of $\psi^\prime$ production at $p_{\rm lab}=6.23$ GeV/c,
the $\bar p$ rescattering acts in a similar way as in the case of $J/\psi$, but the $\psi^\prime$ rescattering has a much larger relative effect
due to a larger $\psi^\prime n$ cross section. The CT effect is also stronger for $\psi^\prime$ production, mostly due to the larger beam momentum.
Note that the CT effect tends to disappear at large transverse momentum of the charmonium. This can be traced back to
the modification of the Sachs electric formfactor of the proton due to CT, as expressed by Eq.(\ref{R}). Thus, with increasing $|t|$ an overall
reduction of the $\bar p n$ scattering amplitude due to $\sigma_{\bar p N}^{\rm eff}$ is partly compensated by the factor $R > 1$. 

\begin{figure}
  \includegraphics[scale = 0.53]{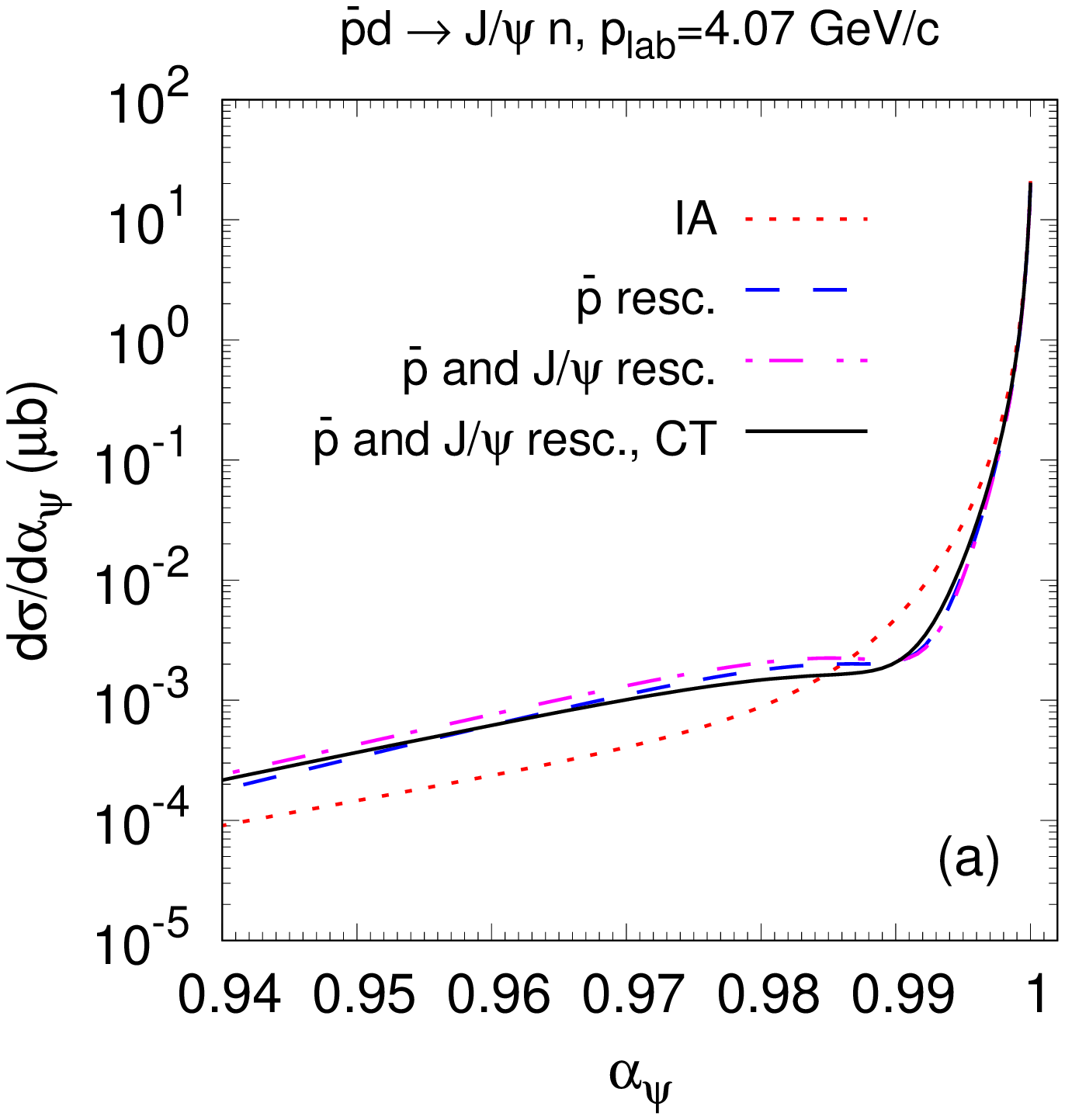}
  \includegraphics[scale = 0.53]{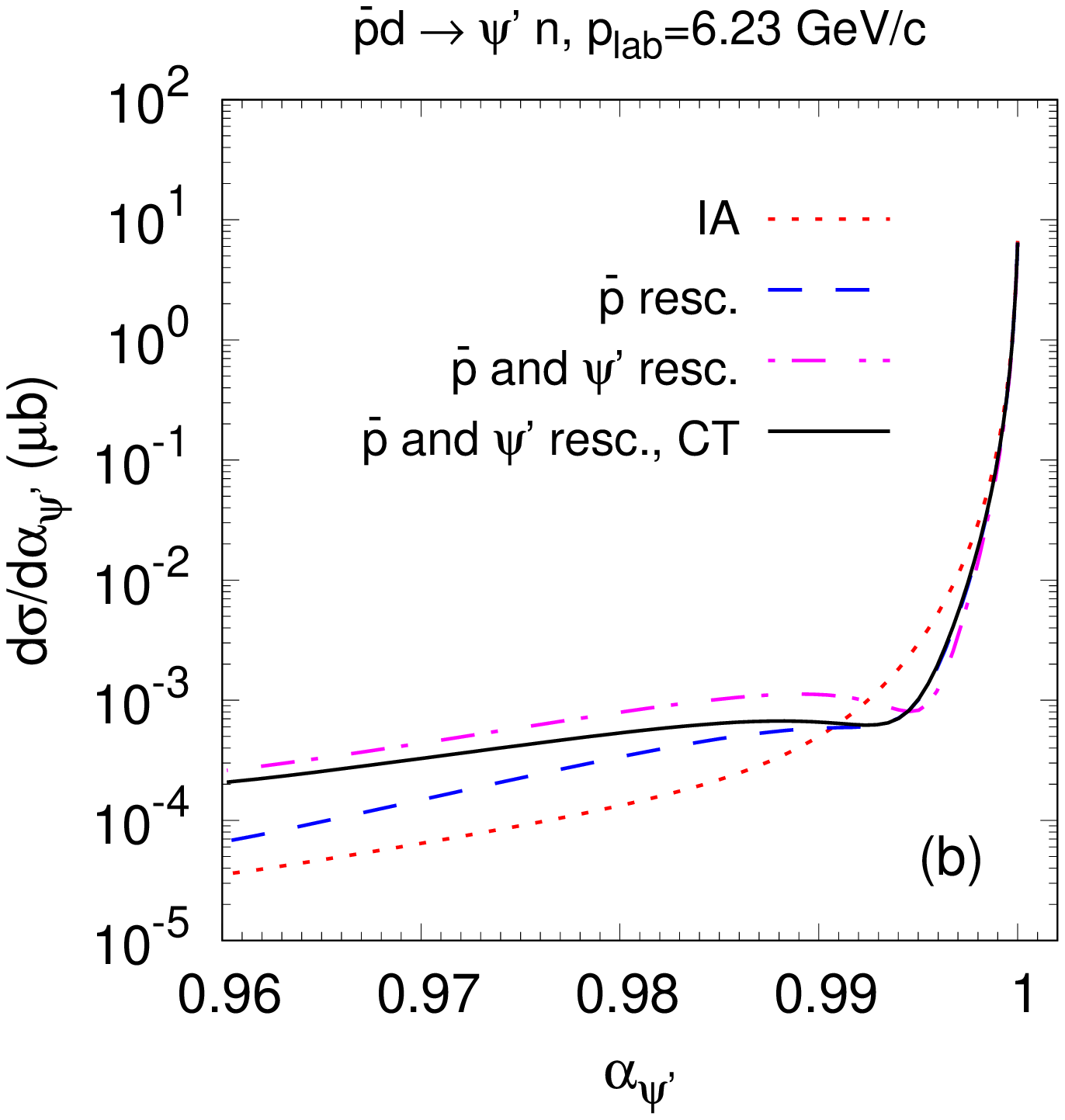}
  \caption{\label{fig:dsigdalpha} Same as in Fig.~\ref{fig:dsigd2pt}, but for the differential cross sections in the light cone momentum
    fraction of charmonium.}
\end{figure}
Fig.~\ref{fig:dsigdalpha} shows the charmonium differential cross section as function of the light cone momentum fraction $\alpha_\psi$ (Eq.(\ref{dsigma_dalpha_psi})).
The lowest values of $\alpha_\psi$ correspond to the largest transverse momenta $p_{\psi t}$. The cross sections are sharply peaked at $\alpha_\psi=1$
reflecting the dominating quasifree mechanism. We see that rescattering helps to decelerate the charmonia.
Since the single elastic rescattering is the leading order correction to the IA diagram, we expect that a similar behavior of
the $\alpha_\psi$-differential cross section should show up also in inclusive charmonium production in $\bar p$-induced
reactions on heavier targets close to the quasifree threshold. 

\begin{figure}
  \includegraphics[scale = 0.53]{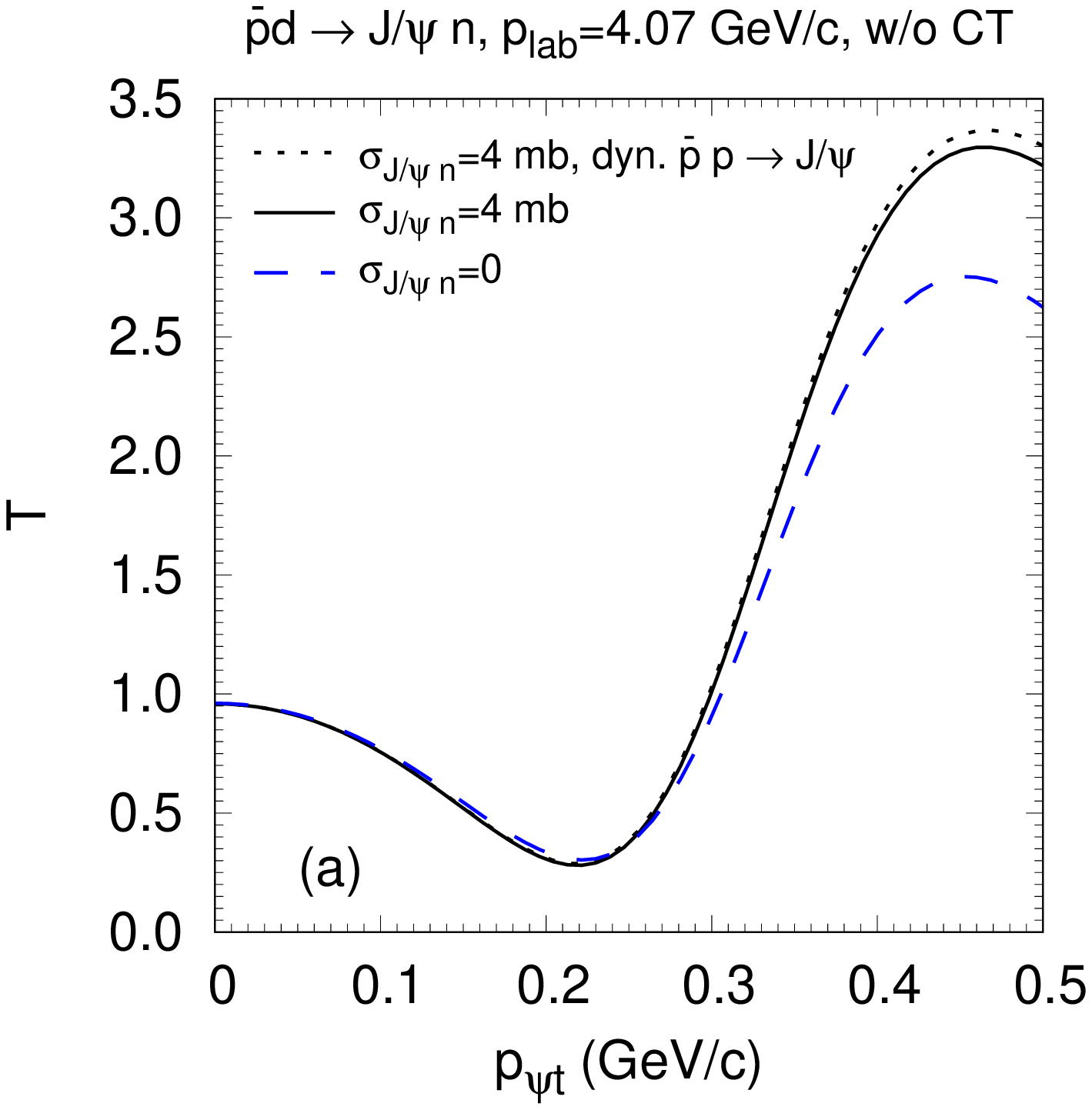}
  \includegraphics[scale = 0.53]{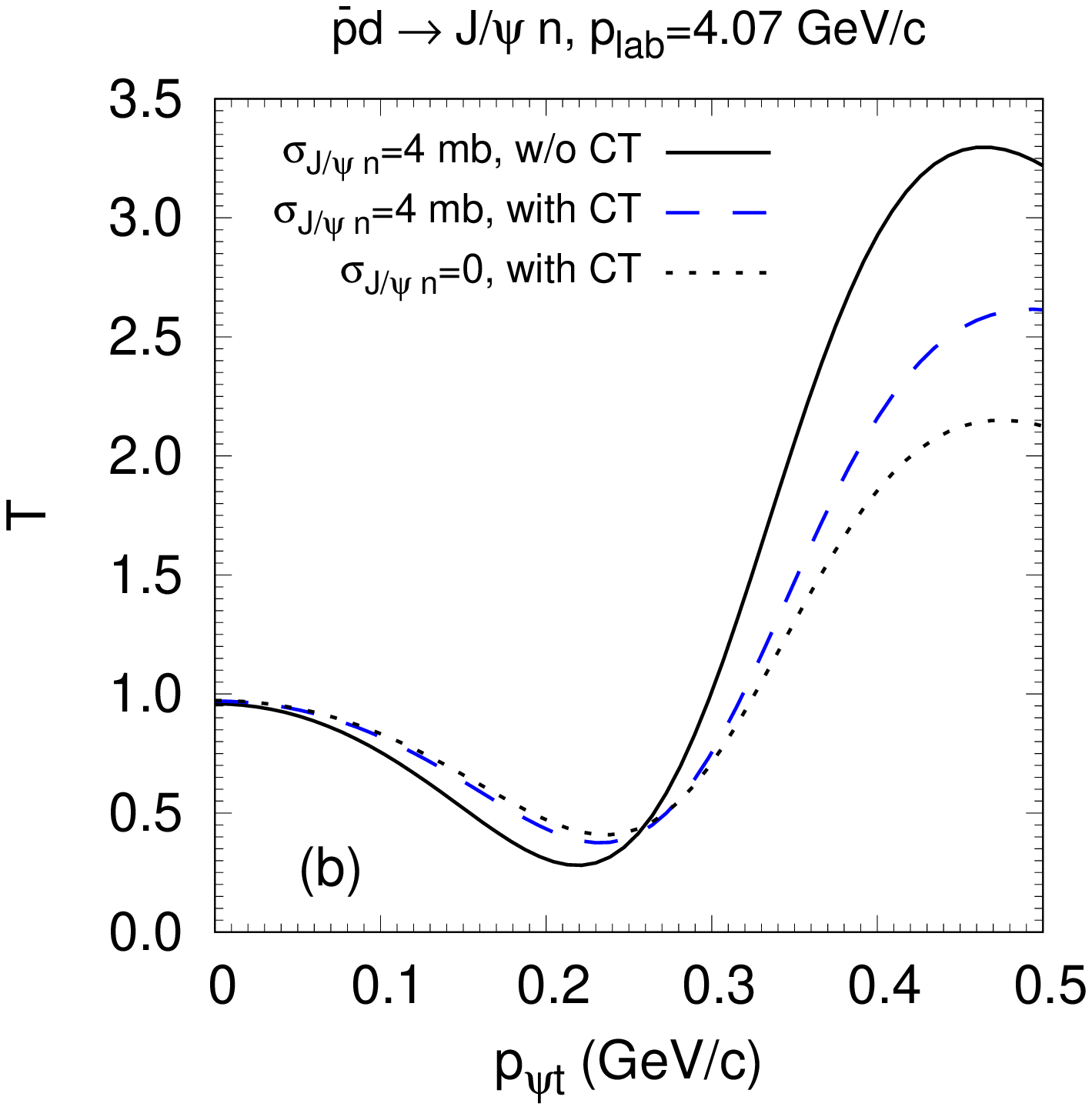}
  \includegraphics[scale = 0.53]{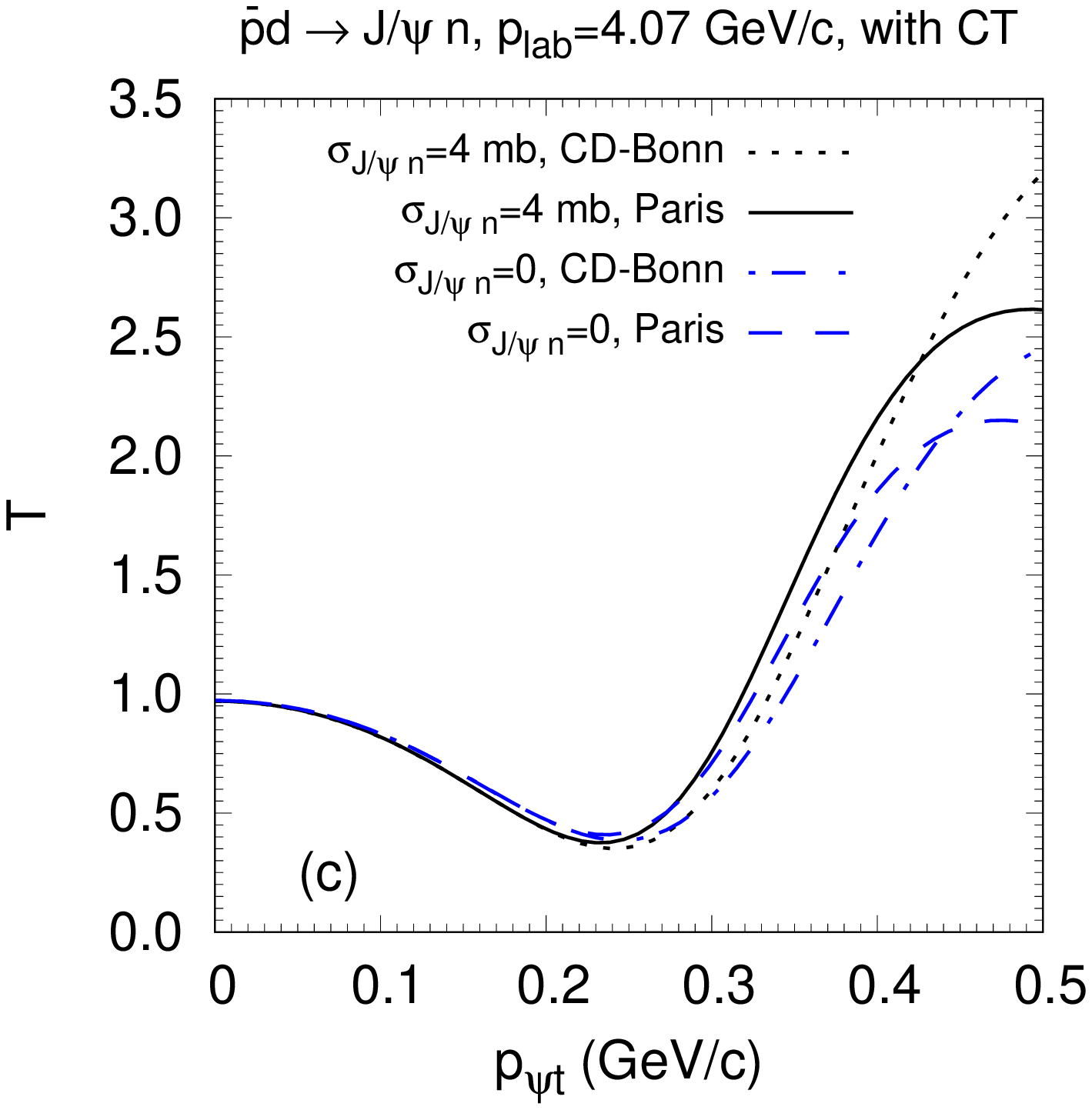}
  \caption{\label{fig:T} Transparency ratio $T$ as a function of the $J/\psi$ transverse momentum for the $\bar p d \to J/\psi\, n$ process
   at $p_{\rm lab}=4.07$ GeV/c.
  Panel (a) shows the results without CT for the Paris DWF: dashed line -- $\bar p$ rescattering,
  solid line -- $\bar p$ and $J/\psi$ rescattering, dotted line -- $\bar p$ and $J/\psi$ rescattering with dynamical treatment of
  the $\bar p p \to J/\psi$ amplitude, i.e. beyond the quasifree approximation.
  Panel (b) shows the results for the Paris DWF: solid (dashed) line -- $\bar p$ and $J/\psi$ rescattering without (with) CT, 
  dotted line -- $\bar p$ rescattering with CT.
  Panel (c) shows the results with CT: dashed (dash-dotted) line -- $\bar p$ rescattering for the Paris (CD-Bonn) DWF,
  solid (dotted) line -- $\bar p$ and $J/\psi$ rescattering for the Paris (CD-Bonn) DWF.}
\end{figure}
The effects of rescattering are better visible in the transverse momentum dependence of the transparency ratio $T$.
As one can see from Fig.~\ref{fig:T}a, the $J/\psi$ transparency ratio varies with the transverse momentum
of the charmonium quite strongly. It first drops from $T \simeq 1$ at $p_{\psi t}=0$  with increasing $p_{\psi t}$
reaching the minimum of $\sim 0.3$ at $p_{\psi t} \sim 0.22$ GeV/c, then starts to increase and reaches the values
of $\sim 3$ at $p_{\psi t} \sim 0.45$ GeV/c. At small $p_{\psi t}$, the IA amplitude dominates because of large
absolute values of the DWF at small proton momenta, while the rescattering amplitudes are relatively small.
With increasing $p_{\psi t}$ the interference terms between the IA amplitude and the rescattering amplitudes lead to the
absorption effect. The rise of the transparency ratio at large transverse momenta is governed by the squared absolute values
of the rescattering amplitudes. This behavior is in-line with earlier calculations of the transparency ratio in the
$d(p,2p)n$ reaction \cite{Frankfurt:1996uz}. $J/\psi$ rescattering leads to $\sim 20\%$ increase of the transparency
ratio at large transverse momenta.

From Fig.~\ref{fig:T}b, we observe that CT reduces the rescattering effects at high transverse momenta by the same amount
the $J/\psi$ rescattering enhances them. At low transverse momenta, CT acts even much more strongly than the $J/\psi$ rescattering
does. Thus, the shape of the transverse momentum dependence of the $J/\psi$ transparency ratio at $p_{\psi t} < 0.2$ GeV/c
is mostly sensitive to the CT effects.

In Fig.~\ref{fig:T}c, we compare the $J/\psi$ transparency ratios calculated with the Paris and CD-Bonn DWF. 
At large transverse momenta the results become uncertain due to the DWF. The calculation with the DWF of CD-Bonn potential
leads to larger $T$-values at large $p_{\psi t}$. This is a consequence of the reduced deuteron momentum distribution at large
momenta for the CD-Bonn DWF, which results in the reduced cross section at large $p_{\psi t}$ in the IA calculation leading
to a relatively larger contribution of the rescattering. However, there is a remarkable independence of $T$ at low transverse
momenta on the DWF: at $p_{\psi t} < 0.2$ GeV/c the calculations with DWFs of Paris and CD-Bonn potentials are almost indistinguishable.

\begin{figure}
  \includegraphics[scale = 0.53]{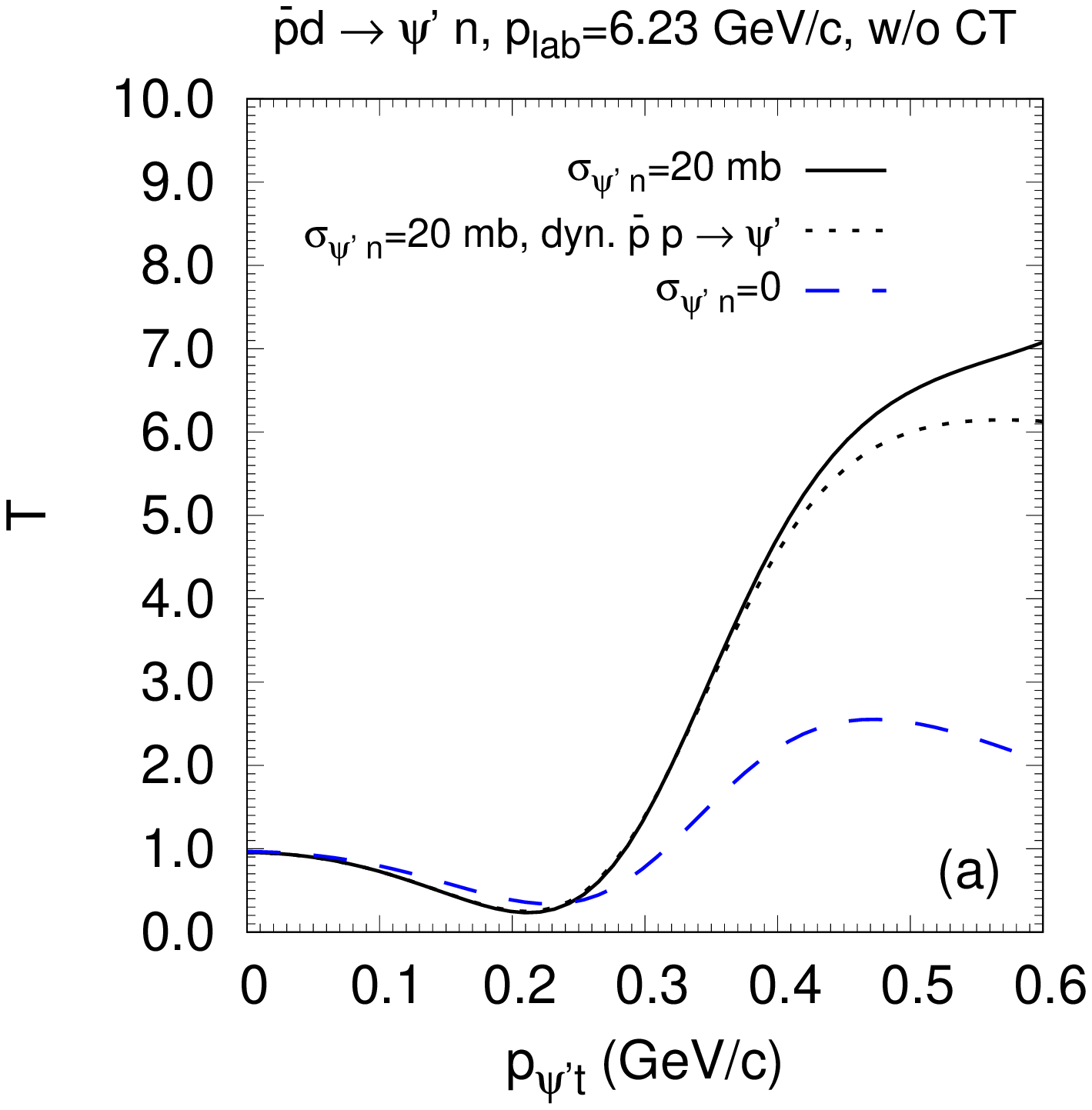}
  \includegraphics[scale = 0.53]{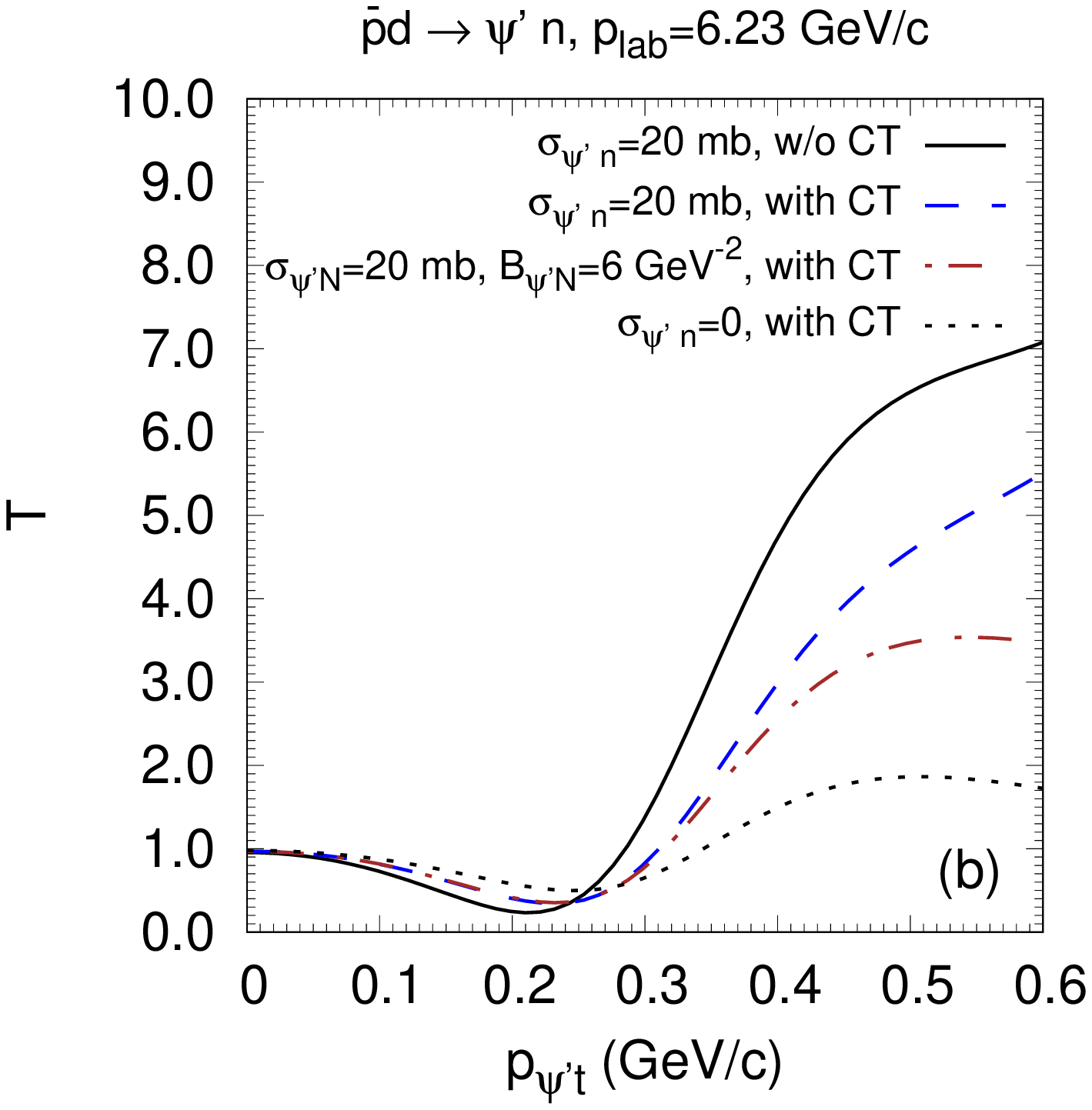}
  \includegraphics[scale = 0.53]{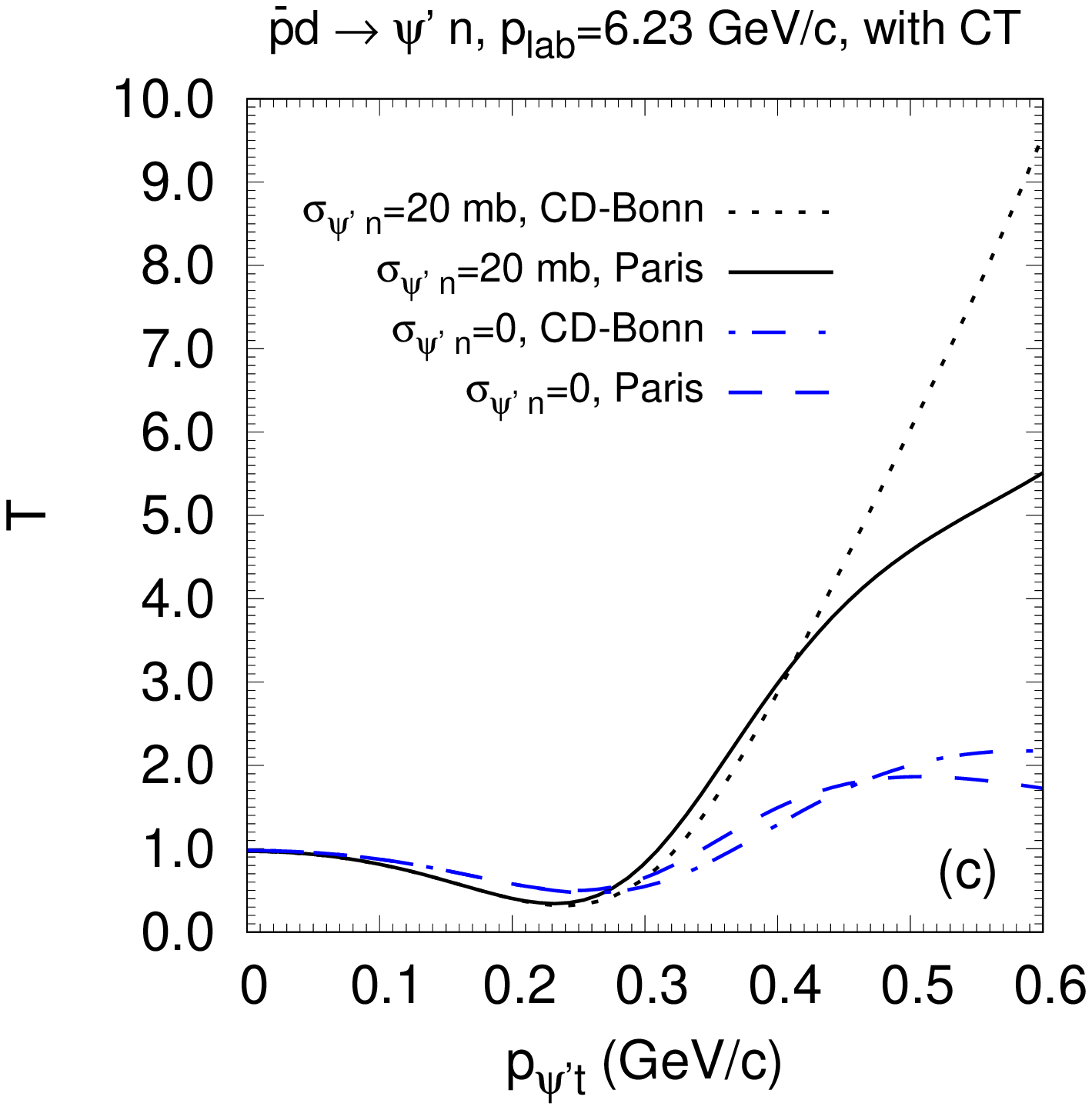}
  \caption{\label{fig:T_Psip} Same as in Fig.~\ref{fig:T}, but for $\psi^\prime$ production at $p_{\rm lab}=6.23$ GeV/c.
  In addition, on the panel (b), the calculation with $B_{\psi^\prime n}=6$ GeV$^{-2}$ with CT is shown by the dash-dotted line.}
\end{figure}
As displayed in Fig.~\ref{fig:T_Psip}a for $\psi^\prime$ production, $\psi^\prime$ rescattering leads to more pronounced
changes in the transparency ratio. At large transverse momenta, $T$ increases by about a factor of three
due to the $\psi^\prime$ rescattering. However, $T$ at high transverse momenta is quite strongly sensitive to CT (Fig.~\ref{fig:T_Psip}b)
and to the DWF (Fig.~\ref{fig:T_Psip}c).
In the absorption region ($p_{\psi^\prime t} < 0.2$ GeV/c), the transparency ratio for $\psi^\prime$ shows a
stronger variation with the charmonium-neutron cross section than in the case $J/\psi$ production. (This is of course not surprising
because the range of expected values of the cross section is larger for $\psi^\prime$.) For example, at $p_{\psi^\prime t} \sim 0.2$ GeV/c
the $T$-value drops by 30\% when $\sigma_{\psi^\prime\,n}$ is increasing from 0 to 20 mb. However, CT has a comparable effect with the variation of the
$\psi^\prime n$ cross section at small transverse momenta. Further modifications have no effect at small $p_{\psi^\prime t}$.
For example, changing the slope parameter of the momentum dependence of the $\psi^\prime n$ cross section $B_{\psi^\prime n}$ from 3 GeV$^{-2}$ (default value)
to 6 GeV$^{-2}$, does not affect the transparency ratio at low transverse momenta.
We have also checked that the transparency ratio is stable with respect to the variation of the ratio of real-to-imaginary parts of the forward scattering amplitude
$\rho_{\psi n}$ in the range $0-0.4$ in the total considered transverse momentum range. 

\begin{figure}
\includegraphics[scale = 0.53]{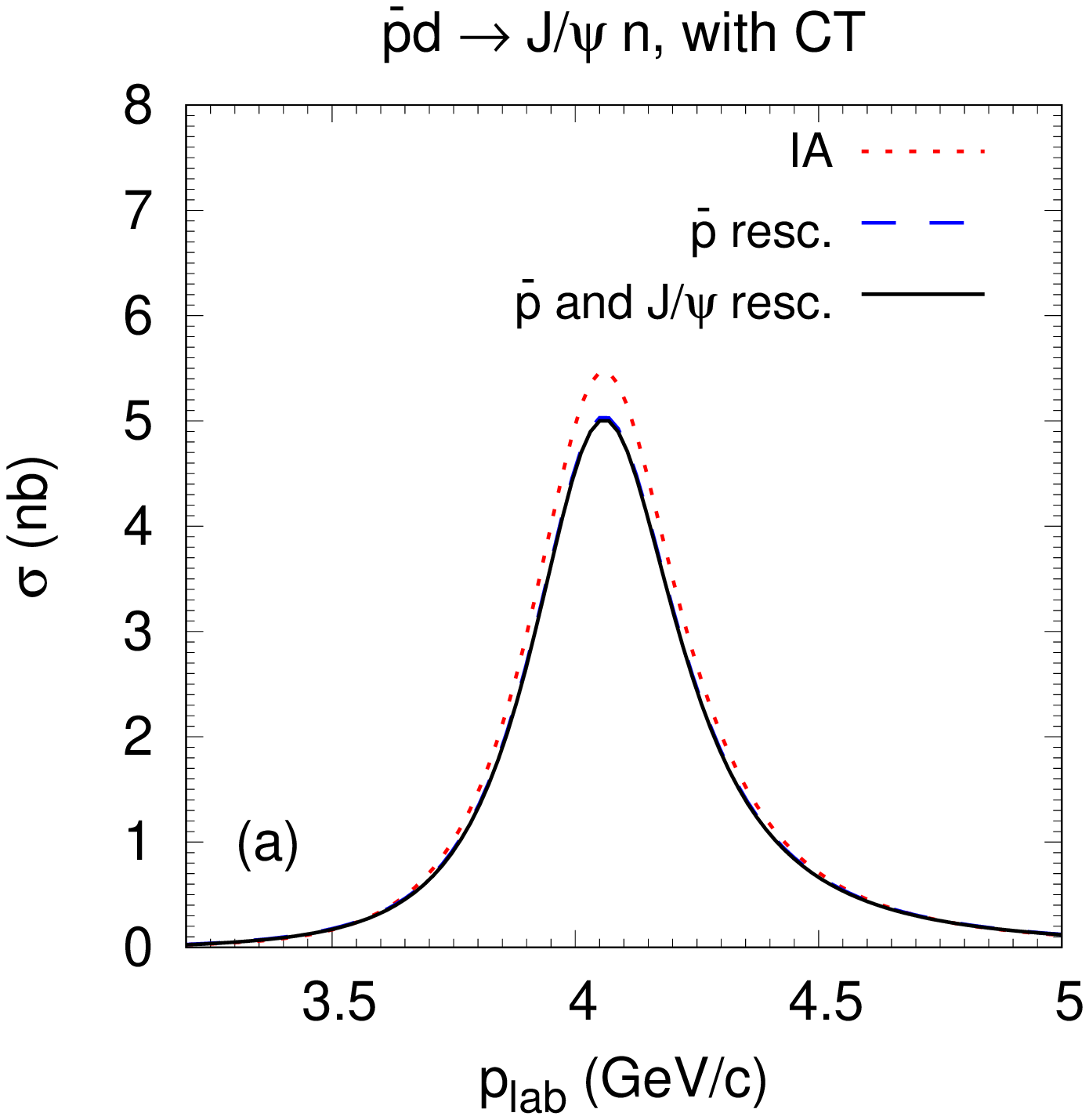}
\includegraphics[scale = 0.53]{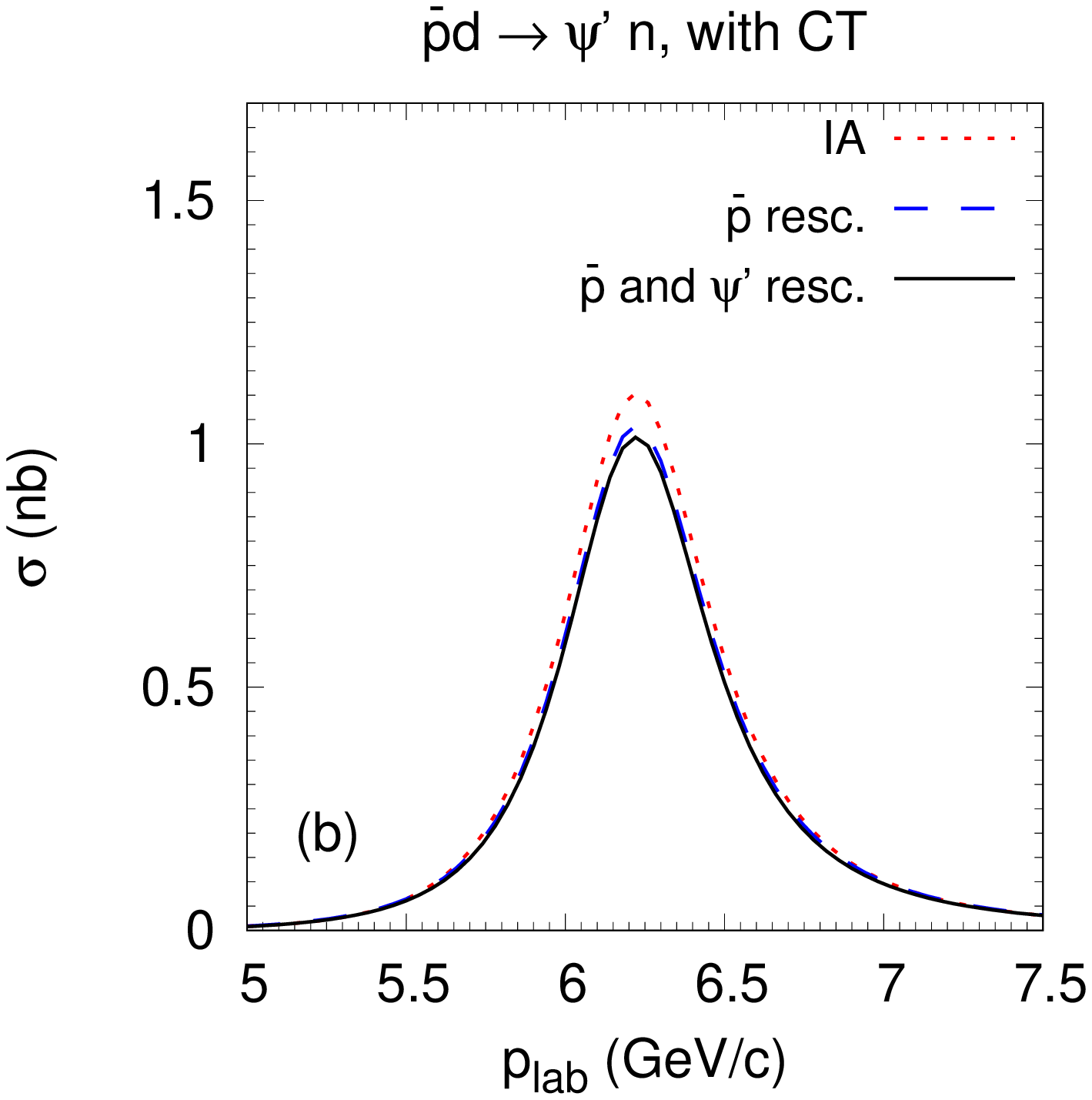}
\caption{\label{fig:sigtot_plabDep} Integrated cross section of charmonium production in $\bar p d \to J/\psi\, n$ (a)
  and $\bar p d \to \psi^\prime\, n$ (b) processes as a function of beam momentum.
  Dotted line -- IA, dashed line -- $\bar p$ rescattering, solid line -- $\bar p$  and charmonium rescattering.
  The rescattering amplitudes are calculated with CT effects.}
\end{figure}
The integrated charmonium production cross sections are shown in Fig.~\ref{fig:sigtot_plabDep}. They are governed by the IA amplitude
at low transverse momenta. Since the rescattering and CT effects are significant only at finite values of the transverse momentum of $\psi$,
where the cross section is small, the integrated cross sections of charmonium production are very weakly sensitive to these effects.
Antiproton rescattering reduces the $J/\psi$ and $\psi^\prime$ production cross sections by $\sim10\%$ at the peak, while the effect
of charmonium rescattering is the reduction of the peak cross section for $J/\psi$ ($\psi^\prime$) by $\sim0.6\%$ ($\sim3\%$). 

\begin{figure}
  \includegraphics[scale = 0.53]{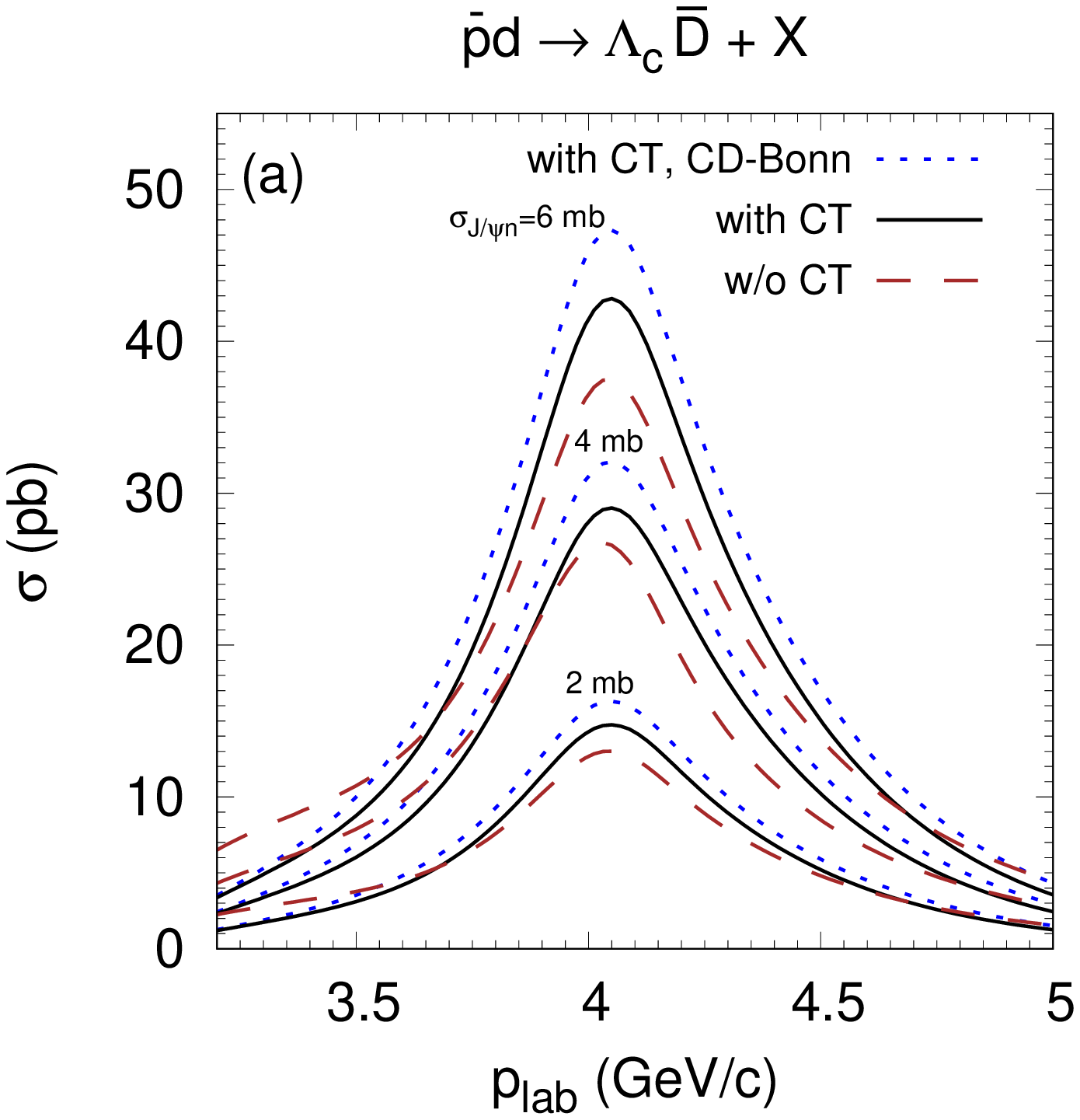}
  \includegraphics[scale = 0.53]{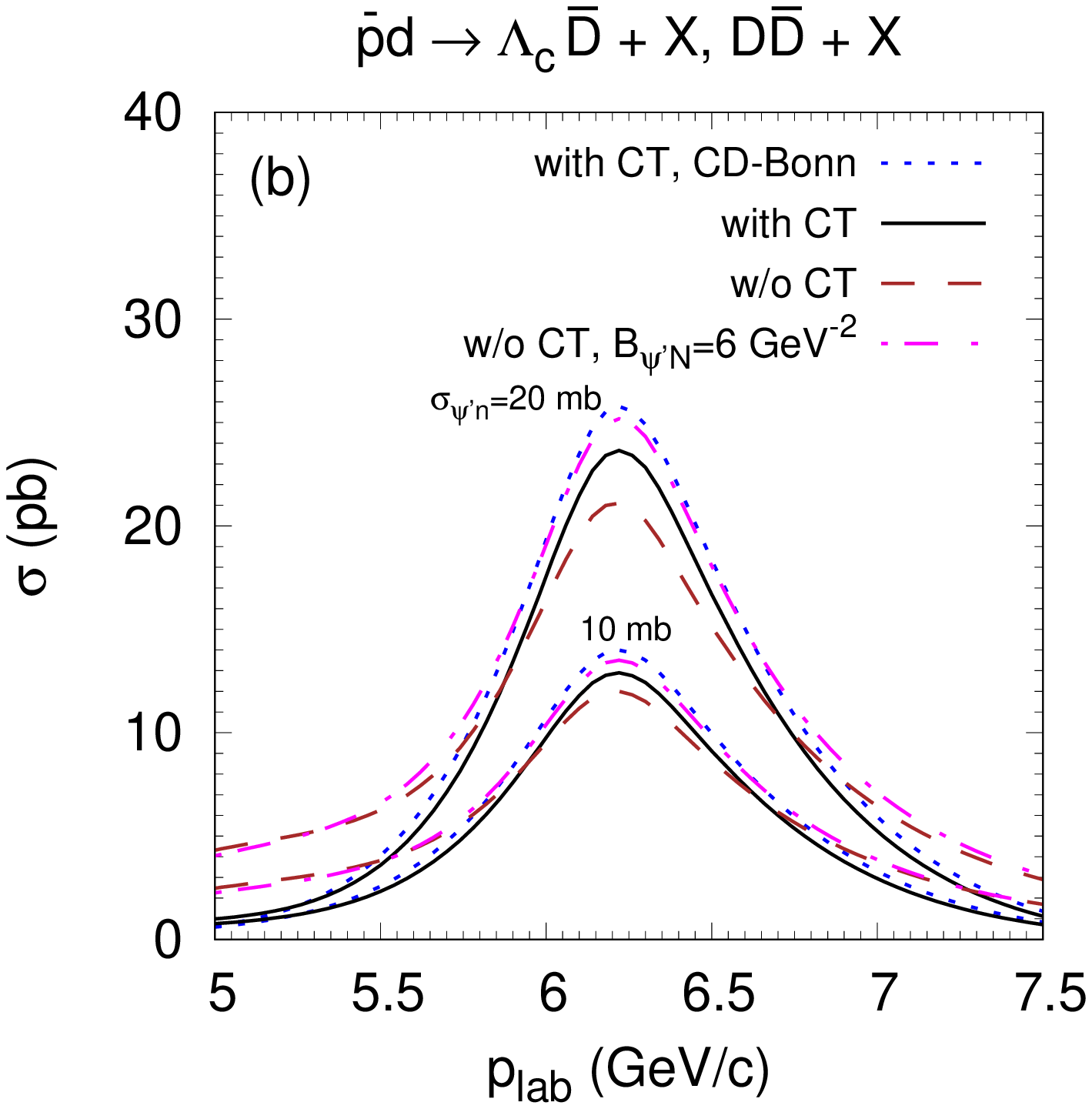}
  \caption{\label{fig:sigDiff_plabDep_varSigma} Inclusive cross section of open charm production in $\bar p d$ collisions vs beam momentum
    near the $J/\psi$ (a) and $\psi^\prime$ (b) production thresholds (see Eqs.(\ref{sigma_LcDbar}),(\ref{sigma_LcDbar_DDbar})).
    The solid (dashed) lines are calculated with (without) CT by using the DWF of the Paris potential.
    The dotted lines are calculated with CT by using the DWF of the CD-Bonn potential.
    The dash-dotted lines on panel (b) are calculated without CT with the value of the slope parameter $B_{\psi^\prime N}=6$ GeV$^{-2}$
    by using the DWF of the Paris potential.
    The set of lines from top to bottom corresponds to decreasing charmonium-neutron total cross section as indicated.}
\end{figure}
On the basis of simple probabilistic arguments, it is possible to evaluate the open charm production cross section in the region of beam momenta
where this process is dominated by the dissociation of $J/\psi$ produced in the quasifree $\bar p p \to J/\psi$ process, i.e. at $p_{\rm lab} \simeq 4.07$ GeV/c.
In this case, the produced $J/\psi$ may only disappear in nuclear strong interactions in reactions $J/\psi\, n \to \Lambda_c \bar D + \mbox{up to 3 pions}$
(see discussion in ref. \cite{Larionov:2013axa}). The threshold for the $\bar p N \to J/\psi\, \pi$ quasifree process is at $p_{\rm lab} =4.55$ GeV/c, and
thus its contribution to the primordial $J/\psi$ production can be neglected. Hence, the inclusive cross section of $\Lambda_c \bar D$ production
in antiproton-deuteron collisions can be calculated as
\begin{equation}
   \sigma_{\bar p d \to \Lambda_c \bar D + X}=\sigma_{\bar p d \to J/\psi\, n}^{w/o\,J/\psi\, resc.}
   - \sigma_{\bar p d \to J/\psi\, n}~,                                       \label{sigma_LcDbar}
\end{equation}
where $\sigma_{\bar p d \to J/\psi\, n}$ is the cross section of $J/\psi$ production calculated with $\bar p$ and $J/\psi$ rescattering contributions,
and $\sigma_{\bar p d \to J/\psi\, n}^{w/o\,J/\psi\, resc.}$ is the same cross section, but calculated without $J/\psi$ rescattering. Eq.(\ref{sigma_LcDbar})
reflects the property of the Glauber theory (cf. refs. \cite{Frankfurt:1996uz,Larionov:2013nga}), that the absorptive corrections
to the modulus squared of the IA amplitude on the nucleus are given by all possible interference terms of the amplitudes with elastic rescattering
in the initial and final states, including the interference between the IA amplitude and any rescattering amplitude.

Similar arguments can be applied near the $\psi^\prime$ threshold, $p_{\rm lab} \simeq 6.23$ GeV/c. The produced $\psi^\prime$ may disappear in
inclusive processes $\psi^\prime n \to \Lambda_c \bar D + X$ and $\psi^\prime n \to N D \bar D + X$. The contribution from the
nondiagonal transition $\psi^\prime n \to J/\psi\, n$, which has a cross section of $\sim 0.1-0.4$ mb \cite{Gerland:2005ca}, can be neglected.
We also neglect the contributions of the processes $\bar p N \to D \bar D$ ($p_{\rm thr}=6.41$ GeV/c) and $\bar p N \to \psi^\prime\, \pi$
($p_{\rm thr}=6.80$ GeV/c). In these approximations, it is possible to extract the open charm production cross section near the $\psi^\prime$
threshold as
\begin{equation}
   \sigma_{\bar p d \to \Lambda_c \bar D + X} + \sigma_{\bar p d \to D \bar D + X}    =\sigma_{\bar p d \to \psi^\prime\, n}^{w/o\,\psi^\prime\, resc.}
   - \sigma_{\bar p d \to \psi^\prime\, n}~,                                       \label{sigma_LcDbar_DDbar}
\end{equation}
where the notations correspond to those in Eq.(\ref{sigma_LcDbar}).

Fig.~\ref{fig:sigDiff_plabDep_varSigma} shows the inclusive open charm production cross section at beam momenta near the quasifree peak for the
$J/\psi$ (a) and $\psi^\prime$ (b). The peak value of the cross section is approximately proportional to the charmonium-neutron total cross section.
The main contribution to the difference of cross sections in Eqs.(\ref{sigma_LcDbar}),(\ref{sigma_LcDbar_DDbar}) is given by the $p_{\psi t}$-integrated
interference term between the IA amplitude and the amplitude with charmonium rescattering. The latter is practically CT-independent. 
Thus, the cross sections of open charm production near the quasifree peaks only weakly depend on the CT effect
(the CT-dependence originates mainly from the interference of the charmonium- and $\bar p$-rescattering amplitudes). 
Since the $p_{\psi t}$-differential cross sections quickly drop with increasing $p_{\psi t}$, the sensitivity of the integrated cross sections
to uncertainties in the DWF at large transverse momenta is quite small. Moreover, also the variation of the slope parameter $B_{\psi^\prime N}$  influences
the open charm cross section at the $\psi^\prime$ threshold rather weakly.

\section{Discussion}
\label{discuss}

Charmonium production in $\bar p d$ collisions can be experimentally studied in the PANDA experiment at FAIR.
Let us now try to estimate the experimental production rates based on our numerical results for the full PANDA design luminosity
$L=2 \cdot 10^{32}~\mbox{cm}^{-2}\mbox{s}^{-1}$ \cite{PANDA}. At the quasifree peaks (see Fig.~\ref{fig:sigtot_plabDep}) thus
in total $5200$ $J/\psi \to e^+ e^-$ events per day and $360$ $\psi^\prime \to J/\psi \pi^+ \pi^- \to e^+ e^- \pi^+ \pi^-$ events per day
are expected to be produced. The rate is effectively doubled (holds for all respective rates given below) if also the
$J/\psi \to \mu^+ \mu^-$ decay mode is detected.    

In the sensitive region for the charmonium-nucleon cross section, i.e. for $0.1~\mbox{GeV/c} < p_{\psi t} < 0.2~\mbox{GeV/c}$, the cross sections
are 366 pb for $J/\psi$ and 58 pb for $\psi^\prime$, as calculated with the default parameters of the elementary amplitudes, including the CT effect.
This would give 380 $J/\psi \to e^+ e^-$ events per day and 20 $\psi^\prime \to J/\psi \pi^+ \pi^- \to e^+ e^- \pi^+ \pi^-$ events per day.
Assuming a measuring time of 14 days, 20\% total event reconstruction efficiency, and adding both dileptonic $J/\psi$ decay modes,
one obtains a statistical accuracy of $\sim 2\%$ for $J/\psi$ and $\sim 9\%$ for $\psi^\prime$.
In the case of $J/\psi$ this would allow to test the presence and
the strength of CT for incoming $\bar p$ (Fig.~\ref{fig:T}b). In the case of $\psi^\prime$, the CT and $\psi^\prime$ absorption are expected to be
of approximately the same strength (Fig.~\ref{fig:T_Psip}b) which complicates the analysis. However, we think that the fit of the transverse
momentum dependence of $T$ (perhaps in a wider $p_{\psi^\prime t}$ region) would allow to disentangle the effects of CT
and of the $\psi^\prime n$ cross section. 

In the case of open charm studies it is sufficient to estimate the production rate of $c$-quark-containing hadrons,
i.e. either $\Lambda_c^+$ or $D$ ($\Sigma_c$ or $D^*$ strong or electromagnetic decays are included in this way too),
since the $\bar c$ quark will be always carried off by $\bar D$.
The $\Lambda_c^+$ production rate in the channel $\Lambda_c^+ \to p K^- \pi^+$ (${\cal B}=6.23\%$ \cite{Tanabashi:2018oca})
is 32 events per day at the $J/\psi$ threshold for $\sigma=30$ pb (see Fig.~\ref{fig:sigDiff_plabDep_varSigma}a).
Assuming that $\psi^\prime$ dissociates on neutron into $\Lambda_c^+ \bar D + X$ would give the $\Lambda_c^+ \to p K^- \pi^+$
production rate of 13 events per day at the $\psi^\prime$ threshold for $\sigma=12$ pb (see Fig.~\ref{fig:sigDiff_plabDep_varSigma}b).
The assumption that $\psi^\prime$ dissociates on the neutron into
$D \bar D + N + X$ with equal probability of $D^+$ and $D^0$ production would give the $D^+ \to K^- 2\pi^+$
(${\cal B}=8.98\pm0.28\%$ \cite{Tanabashi:2018oca}), $D^0 \to K^- \pi^+ \pi^0$ (${\cal B}=14.2\pm0.5\%$ \cite{Tanabashi:2018oca})
and $D^0 \to K^- 2\pi^+ \pi^-$ (${\cal B}=8.11\pm0.15\%$ \cite{Tanabashi:2018oca}) production rates of 9, 15 and 8 events per day, respectively.
Although it may be challenging to identify open charm hadrons in the by many orders larger non-charm hadronic background,
it is worthwhile to study the detection capability of these channels in detail,
due to their strong sensitivity to the charmonium-nucleon cross section.

\section{Summary and conclusions}
\label{summary}

We have theoretically studied exclusive $J/\psi$ and $\psi^\prime$ production in antiproton-deuteron interactions at beam momenta
near the respective quasifree thresholds. The model includes the IA amplitude with the $\bar p p \to \mbox{charmonium}$ transition only,
and the two amplitudes with $\bar p$ and charmonium rescattering on the spectator neutron. The calculations were performed within
the generalized eikonal approximation. CT effects on the $\bar p$ and $\psi$ have been taken into account within the quantum
diffusion model by using the position dependent rescattering amplitudes. The main purpose of the study was to quantify the impact
of the charmonium rescattering on the observables. 

The transverse momentum differential cross sections of charmonium production have a characteristic shape with a depletion at transverse momenta
$p_{\psi t} \ltsim 0.2$ GeV/c and an enhancement at larger transverse momenta. The depletion has an absorptive nature and is due to the interference
of the IA amplitude with the amplitudes with $\bar p$ and $\psi$ rescattering, while the enhancement is due to the squared moduli of the amplitudes
with rescattering and their interference. This behavior is not changed qualitatively by the variation of the model parameters,
although the detailed shapes of the $p_{\psi t}$-dependence of the transparency ratio $T$ differ. The major source of uncertainty in this analysis
is caused by the CT effect for the incoming antiproton which influences its elastic rescattering amplitude on the neutron before charmonium
production.

The region of high charmonium transverse momenta ($p_{\psi t} \gtsim 0.35$ GeV/c) is sensitive to several parameters. Apart from the
values of the $\psi\,n$ cross sections, it is also significantly influenced by the uncertainty in the DWF at large momenta,
by the CT effect for the incoming antiproton, and by the slope parameter $B_{\psi n}$ of the elastic charmonium-neutron amplitude. 

In the region of low transverse momenta ($p_{\psi t} \ltsim 0.2$ GeV/c) the uncertainties due to the DWF and $B_{\psi n}$ disappear.
Still, CT strongly influences the charmonium transparency ratio $T$ in this region of transverse momenta (Fig.~\ref{fig:T}b).
Thus, due to the presumably small $J/\psi\, n$ cross section, $T$ at low $p_{\psi t}$ could be used for the study of CT effects
in the $\bar p d \to J/\psi\, n$ channel. Overall, data on transverse-momentum differential cross sections of the $J/\psi$ and $\psi^\prime$
production could be used to perform multi-parameter fits within our model to simultaneously extract the values of the $\bar p$ coherence length,
charmonium-nucleon cross section, and the slope parameter $B_{\psi n}$.

In the Glauber model, at $p_{\rm lab}$ in vicinity of the quasifree charmonium production threshold, the difference between the integrated
cross sections without and with charmonium rescattering (but always including $\bar p$ rescattering) can be identified with the open charm
meson and baryon production cross section. As this difference is proportional to $\sigma_{\psi\, n}^{\rm tot}$, it can be directly utilized
for the extraction of the charmonium-nucleon cross section (Fig.~\ref{fig:sigDiff_plabDep_varSigma}). The difference is pretty stable
with respect to the inclusion of the CT effect, and to the variation of the DWF and slope parameter $B_{\psi n}$.
Therefore, the data on the cross sections for the channel $\bar p d \to \Lambda_c \bar D + X$ at the $J/\psi$ quasifree threshold and
on the channels $\bar p d \to \Lambda_c \bar D + X$ and $\bar p d \to D \bar D + X$ at the $\psi^\prime$ quasifree threshold could be
used to extract, respectively, the total $J/\psi\, n$ and $\psi^\prime\, n$ cross sections with an accuracy of at least $\sim 30\%$.

The present calculations can be extended to heavier targets where the rescattering effects are stronger.
The study of charmonium production in $\bar p A$ interactions \cite{Larionov:2013axa} revealed the importance
of the detailed behavior of the nucleon densities in the surface region for the quantitative
prediction of $J/\psi$ and $\psi^\prime$ nuclear transparency ratios.
Thus, calculations beyond the local Fermi approximation, taking into
account the shell structure of the target nuclei might lead to new important conclusions.

Lastly, the pentaquark states $P_c(4380)^+$ ($\Gamma=205$ MeV) and $P_c(4450)^+$ ($\Gamma=39$ MeV) \cite{Tanabashi:2018oca}
may also be produced in the two step $\bar p p \to J/\psi, J/\psi p \to P_c^+$ process. The deuteron target allows
to test the pentaquark $P_c^0$ production specifically in the $J/\psi n$ channel, as
molecular interpretations of these states suggest $I=1/2$ \cite{Xiao:2015fia}.

\begin{acknowledgments}
  We thank M.~Bleicher, J.~Haidenbauer, J.~Ritman, and Yu.N.~Uzikov for stimulating discussions. 
  The research of M.S. was supported by the U.S. Department of Energy,
  Office of Science, Office of Nuclear Physics, under Award No. DE-FG02-93ER40771.
  The numerical calculations of the cross sections with the CT effect became possible owing to the computational resources
  of the Frankfurt Center for Scientific Computing (FUCHS-CSC).
\end{acknowledgments}

\bibliography{pbarJpsi}

\end{document}